\documentstyle[tighten,preprint,eqsecnum,aps]{revtex}
\begin{document}

\preprint{IHES/P/95/55}
\title{ The equivalence principle and the Moon}
\author{Thibault Damour}
\address{Institut des Hautes Etudes Scientifiques, 91440
 Bures sur Yvette, France \\
 and D\'epartement d'Astrophysique Relativiste et de
 Cosmologie, Observatoire de Paris, \\
 Centre National de la Recherche Scientifique, 92195
 Meudon CEDEX, France}
\author{David Vokrouhlick\'y\cite{byline1}}
\address{Observatoire de la C\^ote d'Azur, D\'epartement
 CERGA, Avenue N. Copernic, \\ 06130 Grasse, France}
\date{\today}
\maketitle
\begin{abstract}
 The perturbation of the lunar motion caused by a hypothetical
 violation of the equivalence principle is analytically worked
 out in terms of power series \`a la Hill-Brown. The
 interaction with the quadrupolar tide is found to amplify the
 leading order term in the synodic range oscillation by more than
 $62$ \%. Confirming a recent suggestion of Nordtvedt, we find that
 this amplification has a pole singularity for an orbit beyond the
 lunar orbit. This singularity is shown to correspond to the critical
 prograde orbit beyond which, as found by H\'enon, Hill's periodic
 orbit becomes exponentially unstable. It is suggested that ranging
 between prograde and retrograde orbits around outer planets might
 provide future high precision orbital tests of the equivalence
 principle. It is argued that, within the context of string-derived
non-Einsteinian theories, the theoretical significance of orbital
tests of the universality of free fall is to measure the basic
coupling strength of some scalar field through composition-dependent
effects. Present Lunar Laser Ranging data yield the value $\bar{\gamma}
= (-1.2\pm 1.7) \times 10^{-7}$ for the effective Eddington parameter
$\bar{\gamma} \equiv \gamma -1$ measuring this coupling strength.
\end{abstract}

\narrowtext
\section{Introduction}

Gravity seems to enjoy a remarkable universality property: all
bodies are experimentally found to fall with the same acceleration
in an external gravitational field, independently of their mass and
composition. Although Galileo \cite{G1} was the first \cite{G2} to
suggest in a clear and general way that this property of universality
of free fall might hold true, it was left to Newton \cite{N1} to
realize the remarkable conceptual status of this universality:
exact proportionality between a particular force (the weight) and
the general dynamical measure of inertia (the mass). Newton went
further in performing the first precise laboratory tests of the
universality of free fall (pendulum experiments; precision $\sim
10^{-3}$). It is less well known that Newton went even further and
suggested to test the universality of free fall of celestial bodies
by looking for a possible miscentering of the orbits of satellites
around Jupiter, Saturn and the Earth \cite{N2}. More precisely,
Newton considers a possible violation of the ratio weight ($w$)
over mass ($m$), i.e.
\begin{equation}
 {\bar \delta}_{12} \equiv {(w/m)_1 \over (w/m)_2} - 1 \neq 0\; ,
 \label{1one}
\end{equation}
where $1$ labels a satellite and $2$ a planet, the weights $w_1,\,
w_2$ being the gravitational forces exerted by the Sun (label $3$).
He says, without giving any details, that he has found ``by some
computations'' that the center ${\bf x}_c$ of the orbit of the
satellite $1$ around the planet $2$ will be displaced (in the
Sun-planet direction and away from the Sun if ${\bar \delta}_{12}
> 0$) by the amount
\begin{equation}
 \delta |{\bf x}_c -{\bf x}_3| = + {1 \over 2} {\bar
 \delta}_{12} a' \; , \label{1two}
\end{equation}
where $a'$ denotes the radius of the orbit of the planet $2$ around
the Sun. In modern phraseology, one can say that Newton predicted
a ``polarization'' of the satellite's orbit in the Sun-planet
direction (away from the Sun if ${\bar \delta}_{12}> 0$). Then Newton
used his theoretical estimate (\ref{1two}) to conclude from the
observed good centering of the orbits of the satellites of Jupiter
that $|{\bar \delta}_{12}|<10^{-3}$, a number comparable to the
result of his pendulum experiments. Actually, this upper limit
obtained from Jovian satellites is wrong, as Newton's theoretical
estimate (\ref{1two}) is incorrect both in magnitude (being a gross
overestimate in general) and in sign (see below).
We could not find any information about Newton's original
calculations in his published papers. It is surprising that Newton
did not remark that, as a consequence of his estimate (\ref{1two}),
a value $|{\bar \delta}_{12}| = 10^{-3}$ would also entail an
unacceptably large polarization (one fifth) of the Moon's orbit.

As far as we are aware, Laplace was the first to realize that the
best celestial system to test a possible violation of the
universality of free fall (\ref{1one}) is the Earth-Moon system
($1 = {\rm Moon}$, $2 = {\rm Earth}$). In \cite{L1} he derived a
rough estimate of the main observable effect of ${\bar \delta}_{12}$
on the angular motion of the Moon. Then, he noticed that even a
very small ${\bar \delta}_{12}\neq 0$ would spoil the agreement
linking his theoretical derivation of the ``parallactic inequality''
of the Moon, the set of observations of the lunar motion, and the
direct measurements of the solar parallax. He concluded that an
upper bound to the fractional difference in acceleration of the
Moon and the Earth toward the Sun is
\begin{equation}
 |{\bar \delta}_{12}| < {1 \over 3410000} \simeq 2.9 \times 10^{-7}
 \; . \label{1three}
\end{equation}

It is remarkable that the limit (\ref{1three}) is much better than
the limit ($|{\bar \delta}_{AB}| < 2 \times 10^{-5}$) obtained some
years later by Bessel through improved pendulum experiments
\cite{B1}, and has been superseded (though not by much) only by the
work of E\"otv\"os in 1890 ($|{\bar \delta}_{AB}| < 5 \times
10^{-8}$; \cite{E1}). [The later results of E\"otv\"os, Pek\'ar
and Fekete improved the bound to $3 \times 10^{-9}$; \cite{EPF}.] As
we discuss in Appendix~D, in spite of some obscurities in his
reasonings and the lack of a fully accurate calculation of the
effect of ${\bar \delta}_{12}$ in longitude, Laplace's final bound
(\ref{1three}) turns out to be a conservative upper limit, given the
information he had.

In 1907, Einstein \cite{E07} deepened the conceptual
implications of the property of universality of free fall by raising
it to the level of a ``hypothesis of complete physical equivalence''
between a gravitational field and an accelerated system of reference.
This heuristic hypothesis was used very successfully by Einstein in
his construction of the theory of general relativity, and became
later enshrined in the name ``principle of equivalence''.

Within the context of relativistic gravitational theories, the use
of the Moon as a sensitive probe of a possible violation of the
equivalence principle for massive, self-gravitating bodies has been
rediscovered by Nordtvedt in 1968 \cite{N68c}. His idea was that
self-gravitational energies might couple non universally to an
external gravitational field in theories having a different structure
than general relativity \cite{N68a}, \cite{N68b}. [Let us note that
Dicke had mentioned this possibility earlier \cite{D57}, \cite{D61},
\cite{D64} but had not explored its consequences in detail.] Anyway,
Nordtvedt, unaware both of the old ideas of Newton and Laplace, and
of the more recent ones of Dicke, realized that the planned Lunar
Laser Ranging (LLR) experiment was providing an exquisitely sensitive
tool for testing the universality of free fall of massive bodies
\cite{N68c}. Performing a {\it first-order} perturbation analysis
of the lunar orbit (assumed circular and planar) in presence of a
violation of the equivalence principle, ${\bar \delta}_{12} \neq 0$
(see Eq.\ (\ref{1one}) with the labels $1$ and $2$ denoting the
Moon and the Earth, respectively), he provided the first analytical
estimate of the corresponding range oscillation:
\begin{equation}
 \left(\delta r\right)^{(1)} = C^{(1)} {\bar \delta}_{12} a'
 \cos\left[\left(n-n'\right)t + \tau_0\right] \; , \label{1four}
\end{equation}
with
\begin{equation}
 C^{(1)} = {1 + 2n/(n-n') \over n^2 - (n-n')^2} n'^2 \; .
 \label{1five}
\end{equation}
Here, $n$ denotes the (mean) sidereal angular velocity of the Moon
around the Earth, $n'$ the (mean) sidereal angular velocity of the
Earth around the Sun, and $a'$ denotes the radius
of the orbit of the Earth around the Sun (assumed circular). The
angle $\tau \equiv (n-n')t + \tau_0$ is equal to the difference
between the mean longitude of the Moon and the mean longitude of the
Sun (as seen from the Earth). For completeness we derive Eqs.\
(\ref{1four}), (\ref{1five}) in Appendix~A.

Let us note in passing that the $\cos\tau$ dependence of the range
oscillation (\ref{1four}) is equivalent (when disregarding the
perturbations of the motion in longitude) to displacing the center
${\bf x}_c$ of a circular lunar orbit in the Earth-Sun
direction (toward the Sun if ${\bar \delta}_{12}>0$) by the amount
\begin{equation}
 \delta |{\bf x}_c -{\bf x}_3| = - C^{(1)} {\bar
 \delta}_{12} a' \; . \label{1six}
\end{equation}

The result (\ref{1two}) of Newton can therefore be viewed as
implying a range oscillation of the type (\ref{1four}) with
$C^{\rm Newton} = - {1\over 2}$, independently of $n$ and $n'$. By
contrast, the first order estimate (\ref{1five}) contains the small
dimensionless parameter\footnote{ This is the version of the small
parameter which is appropriate to our Hill-Brown treatment. Beware
of the fact that the more traditional perturbation approaches denote
by the letter $m$ the quantity ${\bar m} \equiv n'/n$.}
\begin{equation}
 m = {n' \over n-n'} \; , \label{1seven}
\end{equation}
which is $m \simeq 1/12.3687$ for the Moon and much smaller
for the (Galilean) satellites of Jupiter (e.g. $m \simeq
3.86\times 10^{-3}$ for Jupiter IV). More precisely, Eq.\
(\ref{1five}) can be rewritten as
\begin{equation}
 C^{(1)} = {3\over 2} m {1 + {2 \over 3}m \over 1+{1
 \over 2}m} = {3 \over 2}m \left[1 + {1\over 6}m
 -{1 \over 12}m^2 + \cdots \right] \; . \label{1eight}
\end{equation}

In 1973, Nordtvedt \cite{N73} suggested that a more accurate value
of the coefficient $C$ in the $\cos\tau$ (or ``synodic'') range
oscillation
(\ref{1four}) would be obtained by replacing, in the denominator of
$C^{(1)}$, Eq.\ (\ref{1five}), the angular velocity $n$ by the
frequency of radial perturbations: $n_{\rm rad} = {\dot l} = {\bar
{\rm c}} n$ where the perturbation series giving ${\bar {\rm c}}$
reads (see e.g. \cite{BC}, \cite{BI85})
\begin{equation}
 {\bar {\rm c}} = 1 - {{\dot \varpi} \over n} = 1-{3 \over 4}
  m^2 - {177 \over 32} m^3 - \cdots \; . \label{1nine}
\end{equation}
[In the case of the Moon, the series (\ref{1nine}) is very slowly
convergent. The full value of $1-{\bar {\rm c}} \simeq
0.008572573$ \cite{BC} is more than twice the lowest-order
correction ${3\over 4}m^2$.] The correction of
Ref.~\cite{N73} amounts numerically to increasing the first-order
result (\ref{1five}) by about $13$ \%.

In 1981, Will \cite{W81} tried, more systematically, to estimate
the higher-order corrections in the coefficient $C$ due to
the mixing between the perturbation (\ref{1four}) at frequency
$n-n'$ and the tidal perturbations at frequencies $0$ and $2(n-n')$.
He suggested that the first order result should be multiplied by a
factor $1+2n'/n = 1+2m +{\cal O}(m^2)$, i.e. amplified
by about $15$ \%. As a result of these (coincidentally equivalent)
prescriptions, the literature on the ``Nordtvedt effect'' \cite{W76},
\cite{S76}, \cite{W81}, \cite{D89}, \cite{M91} has, for many years,
used as standard estimate for the range oscillation $\delta r =
C {\bar \delta}_{12} a' \cos\tau$ a value $C \simeq 1.14\, C^{(1)}$
[corresponding to about $9.3\, \eta \cos\tau$ meters in
metrically-coupled theories; see below].

Actually, both modifications (suggested in \cite{N73} and \cite{W81})
of the first-order result fall short of giving an accurate estimate
of the effects due to higher powers of $m$. In fact, they do
not even give correctly the second order in $m$. For
completeness, we compute in Appendix~A, by the standard perturbation
theory of de Pont\'ecoulant \cite{dP}, the contribution at order
${\cal O}(m^2)$ and find that it amounts to multiplying the
first-order result by $1+{9\over 2}m+{\cal O}(m^2)$,
i.e.
\begin{eqnarray}
 C^{(1)}+C^{(2)} &= & \left[1+ {9\over 2}m + {\cal O}\left(
 m^2\right)\right] C^{(1)} \nonumber \\
 &=& {3\over 2}m\left[1+ {14\over 3}m + {\cal O}\left(
 m^2\right)\right]\; . \label{1ten}
\end{eqnarray}

By contrast, \cite{N73} and \cite{W81} would give ${11\over 12}$ and
${13\over 6}$, respectively, for the coefficient of $m$ in
the correcting factor within square brackets in the second Eq.\
(\ref{1ten}). Note that ${14\over 3}m = 37.7$ \% for the Moon.
Recently, Nordtvedt \cite{N95} argued that the full perturbation
series in powers of $m$ (created by the interaction with the
orbit's tidal deformation) will cause a rather large numerical
amplification of the synodic oscillation (\ref{1four}). However,
he gave
only an incomplete theoretical analysis of this amplification. The
only literal result he gives (his Eq.\ (2.33)) is
equivalent to the second-order effect derived in Appendix~A and
leading to Eq.\ (\ref{1ten}). We could not extract from his paper a
precise value for the full effect of the $m$-series.

The aim of the present paper
is to provide, for the first time, a full-fledged Hill-Brown
analytical treatment of the orbital perturbations caused by a
violation of the equivalence principle. Our results will allow
us notably to give a precise numerical value for the full range
oscillation in the case of the actual Moon\footnote{Note, however,
that we consider only the Main Lunar
Problem, i.e. that we neglect the terms proportional to the
squares of the lunar and solar eccentricities, and to the square
of the lunar inclination, which are expected to modify our
numerical estimates by $\lesssim 1\%$.}. Namely, we obtain below
\begin{equation}
 \delta r = 2.9427 \times 10^{12} \bar{\delta}_{12} \cos \tau
\;\; {\rm cm} \; , \label{1eleven}
\end{equation}
corresponding to a full coefficient $C= {3 \over 2} m\times
1.62201$ which is larger than the first-order value (\ref{1eight})
by more than 60\%. More generally, we shall be able to discuss in
detail the dependence on $m$ of the range oscillation: see Eqs.
(\ref{2sixty})--(\ref{2sixtytwo}) and Appendix B. These results
are summarized in Fig.1 below. Our results
confirm the suggestion of Ref.~\cite{N95} that when $m$
increases (corresponding to prograde orbits beyond the actual
lunar orbit) the $\cos\tau$ range oscillation eventually becomes
resonant and
is (formally) infinitely amplified. We have some doubts, however,
about the practical
utility of such a resonant orbit, notably because we show that it
occurs precisely at the value
\begin{equation}
m = m_{cr} = 0.1951039966 \ldots \label{1twelve}
\end{equation}
(corresponding to a sidereal period $T_{cr} = m_{cr} / (1+m_{cr})
\; {\rm yr} = 1.95903 \; {\rm month}$) where the
orbit becomes exponentially unstable. Armed with our theoretical
understanding of the $m$-dependence of the $\cos\tau$ oscillation, we
 suggest other orbits that might be practically interesting
(retrograde orbits, orbits around outer planets). Finally, we emphasize
that within the context of modern unified theories the most probable
theoretical significance of {\it orbital} tests of the universality of
free fall is the same as that of {\it laboratory} tests, namely to
measure, through composition-dependent effects, the strength of the
coupling to matter of some long range scalar field(s). The basic measure
of this coupling strength is embodied in an effective Eddington
parameter $\bar{\gamma} \equiv \gamma-1$ which governs both the
standard post-Newtonian effects (including the violation of the strong
equivalence principle $\propto \eta \equiv 4 \bar{\beta} -\bar{\gamma}$)
and the composition-dependent couplings (violation of the weak
equivalence principle). Actually, string theory suggests that the former
contribution (proportional to the gravitational binding energy)
is negligible compared to the one due to a violation of the weak
equivalence principle. Interpreting the latest LLR observational results
within a recently studied class of string-derived theoretical models,
we conclude that present orbital tests give the excellent constraint
$\bar{\gamma} = (-1.2 \pm 1.7) \times 10^{-7}$. [This limit is comparable
to the (similarly interpreted) constraint coming from laboratory tests
 of the weak equivalence principle: $\bar{\gamma} = (-0.8 \pm 1.0) \times
10^{-7}$.]

The plan of this paper is as follows. Section II presents our
Hill-Brown approach and gives the analytical results obtained with
it. Section III discusses the physical consequences of our results.
Many technical details are relegated to some Appendices: Appendix~A
presents the standard de Pont\'ecoulant treatment of lunar theory
and uses it to derive the second-order result (\ref{1ten}),
Appendix~B gives some details of our Hill-Brown treatment, Appendix~C
treats the link between certain commensurabilities of frequencies,
linear instability and the presence of pole singularities in
perturbed motions, and finally Appendix~D discusses Laplace's
derivation of the remarkably good limit (\ref{1three}) on ${\bar
\delta}_{12}$.

\section{ Hill-Brown treatment of equivalence-principle-violation
 effects}

\subsection{ Introduction}
Relativistic effects in the lunar motion have been investigated by
many authors. The pioneers in this field are de Sitter \cite{dS16}
(who computed the general relativistic contributions to the secular
motions of the lunar perigee and node as observed in a global,
barycentric frame) and Brumberg \cite{B58} (who gave a comprehensive
Hill-Brown treatment of the post-Newtonian 3-body problem). Later
works studied non-Einsteinian effects, notably those associated with the
Eddington post-Newtonian parameters $\beta$ and $\gamma$. The most
comprehensive and accurate analytical study of post-Newtonian
effects in the lunar motion (described in a barycentric frame) is
due to Brumberg and Ivanova \cite{BI85}. For general accounts and
more references see the books \cite{S89} and \cite{B91}. Let us
also mention the semi-analytical treatment of the general
relativistic perturbations of the Moon by Lestrade and
Chapront-Touz\'e \cite{LCT82}.

However, apart from the work of Nordtvedt \cite{N68c}, \cite{N73},
\cite{N95}, the studies of non-Einsteinian effects in the lunar
motion have not considered the effect of a violation of the
equivalence principle. The results of the present paper can therefore
be considered as a completion of Ref.~\cite{BI85} which gave an
accurate Hill-Brown theory of all the other Einsteinian and
non-Einsteinian effects. In fact, as pointed out long ago by
Nordtvedt, the effects of a violation of the universality of free
fall are the most prominent non-Einsteinian effects in the lunar
orbit, and therefore deserve an accurate study. Indeed, most of the
non-Einsteinian effects are {\it non-null} effects, i.e. correspond to
modifications proportional to ${\bar \beta} \equiv \beta-1$ or ${\bar
\gamma}\equiv \gamma-1$ of observable relativistic effects (as seen in
a local, geocentric frame) predicted by Einstein's theory. As the
latter are at the few centimeter level \cite{S86}, \cite{S89},
\cite{SS92}, \cite{N95}, which is the precision of the LLR data,
they can be of no use for measuring ${\bar \beta}$ and ${\bar
\gamma}$ at an interesting level (say $<10^{-2}$). An exception must
be made for secular effects and for the parameters describing the
temporal and spatial transformation linking a local, geocentric
frame to a global, barycentric one, e.g. the parameters entering
the de Sitter-Fokker (``geodetic'') precession. [Recent work
\cite{WND95} concludes that geodetic precession alone constraints
${\bar \gamma}$ at the $1$ \% level].

Besides the ``Nordtvedt effect'' proper (i.e. the effect of ${\bar
\delta}_{12} \neq 0$), that we discuss here, there are some other
{\it null} effects which are more sensitive to ${\bar \beta}$ and ${\bar
\gamma}$ than the non-Einsteinian modifications of non-null
general relativistic effects. A subdominant null effect comes
from the violation of the equivalence principle associated with
the gravitational binding energy of the Earth-Moon system. In lowest
approximation (linear in $m$), it is equivalent (see e.g.
Eq.\ (3.14b) of Ref.~\cite{DEF94}) to replacing ${\bar \delta}_{12}$
by
\begin{equation}
 {\tilde \delta}_{12} = {\bar \delta}_{12} - {1\over 3}\eta\left(
 {na \over c}\right)^2 \; , \label{2one}
\end{equation}
where $a$ denotes the semi-major axis of the lunar orbit and
where $\eta$ denotes, as usual, the combination
\begin{equation}
 \eta \equiv 4{\bar \beta}-{\bar \gamma} = 4\beta-\gamma-3 \; .
 \label{2two}
\end{equation}
In general, ${\bar \delta}_{12}$ is the sum of two physically
independent contributions
\begin{equation}
 {\bar \delta}_{12} = \left({\hat \delta}_1 - {\hat \delta}_2\right)
 + \eta\left({E^{\rm grav}_1 \over m_1c^2} - {E^{\rm grav}_2 \over m_2
 c^2} \right)\; . \label{2three}
\end{equation}
The first contribution ${\hat \delta}_{12} \equiv {\hat \delta}_1 -
{\hat \delta}_2$ is generically expected to be present
because the best motivated
modified theories of gravity violate the ``weak equivalence principle'',
i.e. contain, besides Einstein's universal tensor interaction, some
composition-dependent couplings that make laboratory bodies
fall in a non universal way (see e.g. \cite{LH95}, \cite{DP94}). The second
contribution on the right-hand side of Eq.\ (\ref{2three})
(proportional to $\eta$ like the correction in Eq.\ (\ref{2one}))
contains the gravitational self-energy of the bodies ($A=1,2$)
\begin{equation}
 E^{\rm grav}_A = -(G/2) \int_A\int_A d^3x d^3x' \rho({\bf x})
 \rho({\bf x'})/ |{\bf x}-{\bf x'}| \; , \label{2four}
\end{equation}
and is the one first pointed out by Nordtvedt \cite{N68a},
\cite{N68b}. As indicated by Dicke \cite{D57}, \cite{D61}, it is
present in all gravity theories where the effective, locally measured
gravitational ``constant'' may vary from place to place (see e.g.
Section V B of \cite{LH95}). We shall take as nominal values for the
gravitational self-energies of the Moon (label $1$) and the Earth
(label $2$) the values adopted by Williams, Newhall and Dickey
\cite{WND95}, namely $E^{\rm grav}_1/m_1c^2 = -0.19\times 10^{-10}$,
$E^{\rm grav}_2/m_2c^2 = -4.64\times 10^{-10}$, so that
\begin{equation}
 {E^{\rm grav}_1 \over m_1c^2}- {E^{\rm grav}_2 \over m_2c^2} =
 4.45\times 10^{-10} \; . \label{2five}
\end{equation}
We then find numerically that the modification due to the
gravitational binding energy of the Earth-Moon system, $-{1\over 3}
\eta n^2a^2/c^2 = -{1\over 3}\eta G(m_1+m_2)/ac^2$ in Eq.\
(\ref{2one}), is (to first order) equivalent to decreasing the nominal
value (\ref{2five}) by $-0.039\times 10^{-10}$. This represents
a fractional change of (\ref{2five}) by $-0.87$ \% which is probably
smaller than the uncertainty in the estimate (\ref{2five})
associated with our imperfect knowledge of the internal structures
of the Earth and the Moon. These orders of magnitude
illustrate the fact that the overwhelmingly dominant sensitivity
of the lunar motion to non-Einsteinian effects comes from the terms
proportional to ${\bar \delta}_{12}$ that we concentrate upon in the
following.

\subsection{ Three-body Lagrangian}
The Lagrangian describing the $N$-body problem in the currently best
motivated relativistic theories of gravity, i.e. those where gravity
is mediated both by a tensor field and a scalar field with,
generically, composition-dependent couplings (see e.g. \cite{LH95},
\cite{DP94}), can be written as
\begin{equation}
 L_{{\bar \beta},{\bar \gamma},{\bar \delta}} =  L_{GR} + L_{\bar
 \beta} + L_{\bar \gamma} +  L_{\bar \delta} \; , \label{2six}
\end{equation}
where $L_{GR}$ denotes the general relativistic contribution (in
which one should use an effective value of the gravitational
coupling constant $G$ which incorporates the
composition-independent part of the interaction mediated by the
scalar field),
\begin{equation}
 L_{\bar \gamma} = {1 \over 2c^2} {\bar \gamma} \sum_{A \neq B}
 {G m_A m_B \over r_{AB}} \left({\bf v}_A - {\bf v}_B\right)^2\; ,
 \label{2seven}
\end{equation}
denotes the (non-tensor-like) velocity-dependent part of the two-body scalar
interaction (one-scalaron exchange level),
\begin{equation}
 L_{\bar \beta} = -{1 \over c^2} {\bar \beta} \sum_{B \neq A \neq C}
 {G^2 m_A m_B m_C \over r_{AB} r_{AC}}\; , \label{2eight}
\end{equation}
denotes the modification of the non-linear three-body general
relativistic interaction due to the scalar interaction, and
\begin{equation}
 L_{\bar \delta} = {1 \over 2} \sum_{A \neq B} \left({\bar \delta}_A
 + {\bar \delta}_B\right) {G m_A m_B \over r_{AB}}\; , \label{2nine}
\end{equation}
with
\begin{equation}
 {\bar \delta}_A = {\hat \delta}_A + \eta {E^{\rm grav}_A \over
 m_Ac^2} \; , \label{2ten}
\end{equation}
represents (to lowest order) the combined effect of the
composition-dependent couplings (${\hat \delta}_A \neq 0$; violation
of the ``weak equivalence principle'') and of the Dicke-Nordtvedt
contribution due to the spatial variability of the effective
gravitational coupling constant ($\eta = 4{\bar \beta}-{\bar \gamma}
\neq 0$; violation of the ``strong equivalence principle'').
For a direct, field-theory derivation of $L_{\bar \gamma}$ and
$L_{\bar \beta}$ and the expression of the phenomenological
Eddington parameters ${\bar \gamma}=\gamma-1$ and ${\bar \beta}=
\beta-1$ in terms of the basic coupling parameters of the scalar
field (as well as the generalization of these results to the case
of strongly self-gravitating bodies) see Ref.~\cite{DEF92}.

We assume here that all the general relativistic contributions to the
lunar motion (and to its observation through laser ranging) are
separately worked out with sufficient accuracy, using, for instance,
the new, complete framework for relativistic celestial mechanics
of Ref.~\cite{DSX} (which provides the first consistent relativistic
description of the multipole moments of extended bodies). Following
the discussion above, we henceforth discard the subdominant
contributions coming from $L_{\bar \beta}$ and $L_{\bar \gamma}$ to
concentrate upon the effects due to $L_{\bar \delta}$. [The
barycentric frame contributions of $L_{\bar \beta}$ and $L_{\bar
\gamma}$ have been accurately computed by Brumberg and Ivanova
\cite{BI85} and can be linearly superposed with the ones of $L_{\bar
\delta}$.] Finally, it is enough to consider the sum of the
lowest-order approximation to $L_{GR}$ and of $L_{\bar \delta}$,
namely
\begin{equation}
 L({\bf x}_A,{\bf v}_A) = \sum_A {1\over 2}m_A {\bf v}_A^2 + {1\over
 2}\sum_{A\neq B} {G_{AB} m_A m_B \over r_{AB}} \; ,
 \label{2eleven}
\end{equation}
where
\begin{equation}
 G_{AB} = G \left[ 1+{\bar \delta}_A +{\bar \delta}_B\right]\; ,
 \label{2twelve}
\end{equation}
with ${\bar \delta}_A$ of the form given in (\ref{2ten}), represents
the effective, composition-dependent gravitational coupling between
the massive bodies $A$ and $B$. In Eq.\ (\ref{2eleven}) ${\bf v}_A =
d{\bf x}_A/dt$ denotes the (barycentric) velocity of body $A$, $m_A$
the (inertial) mass of $A$ and $r_{AB} = |{\bf x}_A-{\bf x}_B|$ the
distance between $A$ and $B$. We consider a three-body problem, and
more particularly the Moon-Earth-Sun system ($1={\rm Moon}$, $2={\rm
Earth}$, $3={\rm Sun}$). Evidently, all our analytical results will
apply if $2$ denotes another planet, and $1$ one of its natural or
artificial satellites. [Note, however, that the relative orders of
magnitude of the non-Einsteinian effects is different for low-orbit
artificial Earth satellites. See \cite{DEF94} for a recent
discussion.]

Starting from Eq.\ (\ref{2eleven}), we first separate the variables
describing the motion of the center of mass of the Earth-Moon system,
\begin{eqnarray}
 m_0 {\bf x}_0 &\equiv& m_1 {\bf x}_1 + m_2 {\bf x}_2 \; ,
 \label{2thirteen} \\
 m_0 &\equiv& m_1 + m_2 \; , \label{2fourteen}
\end{eqnarray}
and ${\bf v}_0 \equiv d{\bf x}_0/dt$, from those describing the
relative lunar motion,
\begin{eqnarray}
 {\bf x}_{12} &\equiv& {\bf x}_1 - {\bf x}_2 \; , \label{2fifteen} \\
 \mu_{12} &\equiv& {m_1 m_2 \over m_1+m_2}\; , \label{2sixteen}
\end{eqnarray}
and ${\bf v}_{12} \equiv d{\bf x}_{12}/dt$. This yields
\widetext
\begin{equation}
 L({\bf x}_0,{\bf x}_{12},{\bf x}_3;{\bf v}_0,{\bf v}_{12},{\bf v}_3)
  =  {1\over 2}m_0 {\bf v}_0^2 + {1\over 2}\mu_{12} {\bf v}_{12}^2
 + {1\over 2}m_3 {\bf v}_3^2
 + G_{12} {m_1 m_2 \over r_{12}} + G_{13} {m_1 m_3 \over r_{13}}
 + G_{23} {m_2 m_3 \over r_{23}} \; , \label{2seventeen}
\end{equation}
where
\begin{mathletters}
 \label{2eighteen}
\begin{eqnarray}
 r_{13} & = & |{\bf x}_3 - {\bf x}_0 - X_2 {\bf x}_{12}| \; ,
 \label{2eighteena} \\
 r_{23} & = & |{\bf x}_3 - {\bf x}_0 + X_1 {\bf x}_{12}| \; ,
 \label{2eighteenb} \\
 X_1 &\equiv& m_1/m_0\; ,\quad X_2 \equiv m_2/m_0 = 1-X_1\; .
 \label{2eighteenc}
\end{eqnarray}
\end{mathletters}
Expanding $r_{13}^{-1}$ and $r_{23}^{-1}$ in powers of $r_{12}/
r_{30}$ ($r_{30}=|{\bf x}_{30}|$ with ${\bf x}_{30} = {\bf x}_3 -
{\bf x}_0$) leads to
\begin{eqnarray}
 L &=& L_{03}({\bf x}_0,{\bf x}_3,{\bf v}_0,{\bf v}_3) + \mu_{12}
 {\hat L}_{12}({\bf x}_{12},{\bf v}_{12},{\bf x}_{30}) \; ,
 \label{2nineteen} \\
 L_{03} &=& {1\over 2}m_0 {\bf v}_0^2 + {1\over 2}m_3 {\bf v}_3^2 +
 \left(X_1 G_{13} + X_2 G_{23}\right) {m_0 m_3 \over r_{03}} \; ,
 \label{2twenty}  \\
 {\hat L}_{12} &=& {1\over 2} {\bf v}_{12}^2 + G_{12} {m_0 \over
 r_{12}} + R_1 + R_2 + R_3 + \cdots \; , \label{2twentyone}
\end{eqnarray}
\begin{mathletters}
 \label{2twentytwo}
\begin{eqnarray}
 R_1 &=& m_3 \left(-G_{13} + G_{23}\right) x_{12}^i\, \partial^{(3)}_i
 {1 \over r_{30}} \; ,  \label{2twentytwoa}  \\
 R_2 &=& {1 \over 2!} m_3 \left(G_{13}X_2 + G_{23}X_1\right) x_{12}^i
 x_{12}^j\, \partial^{(3)}_{ij} {1 \over r_{30}} \; ,
 \label{2twentytwob}  \\
 R_3 &=& {1 \over 3!} m_3 \left(-G_{13}X_2^2 + G_{23}X_1^2\right)
 x_{12}^i x_{12}^j x_{12}^k\, \partial^{(3)}_{ijk} {1 \over r_{30}}
 \; , \label{2twentytwoc}
\end{eqnarray}
\end{mathletters}
\narrowtext
\noindent where $\partial^{(3)}_i \equiv \partial/\partial x^i_3$,
$\partial^{(3)}_{ij} \equiv \partial^2/\partial x^i_3\partial x^j_3$,
$\ldots$. Note that the suffices $1,2,3,\ldots$ in $R_n$'s have
nothing to do with the body labels $A,B = 1,2,3$, but keep track of
the successive powers of ${\bf x}_{12}$.

To a very good approximation we can consider that, in the (normalized)
Earth-Moon Lagrangian (\ref{2twentyone}), the motion of the Sun with
respect to the Earth-Moon barycenter, ${\bf x}_{30}(t)$, is obtained
by solving the two-body Lagrangian (\ref{2twenty}). After
separating in Eq.\ (\ref{2twenty}) the motion of the center of mass
of the Earth-Moon-Sun system, the reduced Lagrangian describing the
dynamics of the relative motion ${\bf x}_{30}$ is
\begin{equation}
 {\hat L}^{\rm relative}_{03} = {1\over 2}{\bf v}_{03}^2 + G_{03}
 {m_0+m_3 \over r_{03}}\; , \label{2twentythree}
\end{equation}
where we introduced
\begin{equation}
 G_{03} \equiv X_1 G_{13} + X_2 G_{23}\; . \label{2twentyfour}
\end{equation}
Therefore, seen from the Earth-Moon barycenter, the Sun undergoes
a Keplerian, elliptic motion corresponding to a total effective
gravitational mass $G_{03}(m_0+m_3)=G_{03}(m_1+m_2+m_3)$. If we
denote (as is traditional in lunar theory) the mean angular
velocity and the semi-major axis of this elliptic motion as $n'$ and
$a'$, respectively, we can write
\begin{equation}
 n'^2 a'^3 = G_{03}(m_0+m_3)=G_{03}(m_1+m_2+m_3)\; .
 \label{2twentyfive}
\end{equation}
For simplicity's sake, we shall in fact consider the ``Main Problem''
of lunar theory in which the Sun is considered as moving in a
circle of radius $a'$ with the constant angular velocity $n'$.

Evaluating the derivatives with respect to ${\bf x}_3$ in Eqs.\
(\ref{2twentytwo}), and using Eq.\ (\ref{2twelve}), yields
\begin{mathletters}
 \label{2twentysix}
\begin{eqnarray}
 R_1 &=& m_3^{\rm grav} {\bar \delta}_{12} {{\bf N}.{\bf r} \over
 r'^2} \; ,  \label{2twentysixa} \\
 R_2 &=& m_3^{\rm grav} \left[1+\left(X_2-X_1\right){\bar
 \delta}_{12}\right] {3\left({\bf N}.{\bf r}\right)^2 - {\bf r}^2
 \over 2r'^3} \; ,  \label{2twentysixb} \\
 R_3 &\simeq& m_3^{\rm grav} \left(X_2-X_1\right){5\left({\bf N}.
 {\bf r}\right)^3 - 3\left({\bf N}.{\bf r}\right){\bf r}^2 \over
 2r'^4} \; .  \label{2twentysixc}
\end{eqnarray}
\end{mathletters}
To simplify the notation, we have written ${\bf r}\equiv {\bf x}_{12}
={\bf x}_1 - {\bf x}_2$, ${\bf N}\equiv {\bf x}_{30}/r_{30}$
(directed toward the Sun), $r'\equiv r_{30}$, $m_3^{\rm grav}\equiv
G_{03}m_3$, where we recall that $G_{03}$ is the weighted average
(\ref{2twentyfour}), and
\begin{equation}
 {\bar \delta}_{12} \equiv {\bar \delta}_1 - {\bar \delta}_2\; ,
 \label{2twentyseven}
\end{equation}
which agrees with our previous expressions (\ref{1one}) or
(\ref{2three}). In the (much smaller) term $R_3$ we have neglected
the ${\bar \delta}$-modifications.

\subsection{ Hill's equations of motion}
Following Euler's second lunar theory and Hill \cite{Hill},
\cite{Brown}, it is convenient to refer the motion of the Moon to
axes rotating with the mean angular velocity of the Sun. For
simplicity, we shall consider the main lunar problem in which (i)
the Sun is considered as moving in a circle of radius $a'$ with
the uniform angular velocity $n'$, (ii) the Moon moves in the
same plane as the Sun, and (iii) one looks for a periodic motion
of the Moon in the frame rotating with the angular velocity $n'$.
Taking into account the lunar and solar eccentricities $e$ and
$e'$, and the lunar inclination $I$, is expected to modify our
results by terms of order $O (e^2,e'^2,\sin^2 I) \lesssim 1\%$.

With respect to the rotating frame $({\bf e}_X(t),{\bf e}_Y(t))$,
with ${\bf e}_X={\bf N}$ directed toward the Sun, the position
and velocity vectors of the Moon read ${\bf r}={\bf x}_{12}=X
{\bf e}_X + Y {\bf e}_Y$, ${\bf v}={\bf v}_{12}=({\dot X}-n'Y)
{\bf e}_X +({\dot Y}+n'X){\bf e}_Y$ (the overdot denoting $d/dt$).
When expanding the kinetic terms in the reduced, relative Lagrangian
(\ref{2twentyone}), i.e. ${1\over 2}{\bf v}^2= {1\over 2}({\dot
X}-n'Y)^2+{1\over 2}({\dot Y}+n'X)^2$, one recognizes the usual
Coriolis (terms linear in ${\dot X}$ and ${\dot Y}$) and centrifugal
(terms quadratic in $X$ and $Y$) effects. The centrifugal terms can
be gathered with the contribution $R_2$, Eq.\ (\ref{2twentysixb}),
which is also quadratic in $X$ and $Y$. The resulting Lagrangian
describing the dynamics in the rotating frame reads
\widetext
\begin{equation}
 {\hat L}_{12}(X,Y,{\dot X},{\dot Y}) = {1\over 2}\left({\dot X}^2+
 {\dot Y}^2\right) + n'\left(X{\dot Y}-Y{\dot X}\right) + F(X,Y)\; .
 \label{2twentyeight}
\end{equation}
Here, $F(X,Y)=G_{12}m_0/r + R_1 + [R_2 + {1\over 2}n'^2(X^2+Y^2)]+
R_3+\cdots = F_0+F_1+F_2+F_3+\cdots$ is a {\it time-independent}
potential with
\begin{mathletters}
 \label{2twentynine}
\begin{eqnarray}
 F_0 &=& {G_{12} m_0 \over \sqrt{X^2+Y^2}} \; ,  \label{2twentyninea} \\
 F_1 &=& {n'^2 a' \over 1+ m_0/m_3} {\bar \delta}_{12} X\; ,
 \label{2twentynineb} \\
 F_2 &=& {1\over 2}{n'^2 \over 1+ m_0/m_3}\left\{\left[1+\left(X_2-
 X_1\right) {\bar \delta}_{12}\right]\left(3 X^2\right)+\left[{m_0
 \over m_3} -\left(X_2- X_1\right) {\bar \delta}_{12}\right]\left(
 X^2+Y^2\right)\right\}\; ,  \label{2twentyninec} \\
 F_3 &\simeq& {1\over 2}{n'^2 \over a'}\left(X_2-X_1\right) \left[5X^3 -
 3X\left(X^2+Y^2\right)\right]\; . \label{2twentynined}
\end{eqnarray}
\end{mathletters}
\narrowtext

In writing out Eqs.\ (\ref{2twentynine}) we have replaced $m^{\rm
 grav}_3=G_{03}m_3$ by the expression $m^{\rm grav}_3=n'^2a'^3 /(1+
m_0/m_3)$ obtained from Eq.\ (\ref{2twentyfive}). The equations of
motion corresponding to the Lagrangian (\ref{2twentyeight}) read
\begin{mathletters}
 \label{2thirty}
\begin{eqnarray}
 {d^2 X \over dt^2} - 2 n' {dY \over dt} &=& {\partial F \over
 \partial X} \; ,  \label{2thirtya} \\
 {d^2 Y \over dt^2} + 2 n' {dX \over dt} &=& {\partial F \over
 \partial Y} \; .  \label{2thirtyb}
\end{eqnarray}
\end{mathletters}

We see from Eqs.\ (\ref{2twentynine}) that a violation of the
equivalence principle has several consequences in the lunar theory:
(i) the effective gravitational constant appearing in front of the
Earth-Moon total mass $m_0=m_1+m_2$, namely $G_{12}=G(1+{\bar
\delta}_1 + {\bar \delta}_2)$, differs from the one appearing in
the theory of the Earth orbital motion, $G_{03}$ defined in Eq.\
(\ref{2twentyfour}); (ii) there is a new term, linear in $X$, in the
Lagrangian, $F_1$, Eq.\ (\ref{2twentynineb}); (iii) the usual tidal
plus centrifugal potential $F_2$ (as well as the higher-order tidal
potentials) is fractionally modified by ${\bar \delta}_{12} \neq 0$.
The effect (i) has practically no observational consequences as,
for instance, the ``GM'' of the Earth is measured much more
accurately from Earth satellites (artificial or natural) than from
the correction it brings in the Earth-Sun interaction. The effect
(ii) is the one discussed by Newton, Laplace and Nordtvedt, that we
shall study in
detail below. As for the effects (iii) it will be clear from the following
that they are numerically negligible compared to the effects of
$F_1$ because the corresponding source terms in the equations of
motion are smaller by a factor $r/a' \simeq 1/400$, and, moreover,
the corresponding solution is not amplified (as the $F_1$-effects)
by a small divisor $1/m \simeq 12$ because they correspond to
the driving frequency $2(n-n')$ (instead of $n-n'$ for $F_1$).
Finally, as we can also neglect the fractional correction $(1+m_0/
m_3)^{-1}$ to ${\bar \delta}_{12}$ in (\ref{2twentynineb}) [$m_0/
m_3 \simeq 1/328900$], we shall keep Eqs.\ (\ref{2twentyninea}) and
(\ref{2twentynined}) and replace (\ref{2twentynineb}),
(\ref{2twentyninec}) by
\begin{mathletters}
 \label{2thirtyone}
\begin{eqnarray}
 F_1 &\simeq& n'^2 a'{\bar \delta}_{12} X \; ,\label{2thirtyonea} \\
 F_2 &\simeq& {3\over 2}n'^2 X^2 \; . \label{2thirtyoneb}
\end{eqnarray}
\end{mathletters}

The contribution $F_3$ to the potential $F$ (octupolar tide)
generates the so-called ``parallactic'' terms in the lunar motion.
Compared to the usual (quadrupolar) tidal contribution, they
contain the small parameter $r/a' \simeq 1/400$. Hill's approach
consists of solving first exactly (in the sense of infinite power
series) the dynamics defined by the truncated potential $F_{\rm Hill}
=F_0+F_2 = G_{12}m_0/r + {3\over 2} n'^2 X^2$. [The parallactic terms
are obtained later by perturbing Hill's main problem.] In presence
of a violation of the equivalence problem we have to add the term
$F_1$, Eq.\ (\ref{2thirtyonea}), to $F_{\rm Hill}$ (``dipolar tide''!).
The resulting equations of motion (\ref{2thirty}) read explicitly
\begin{mathletters}
 \label{2thirtytwo}
\begin{eqnarray}
 {d^2 X \over dt^2} - 2 n' {dY \over dt} &=& -{G_{12} m_0 \over r^3}
 X \nonumber \\ & & + 3n'^2 X + n'^2 a' {\bar \delta}_{12}
 \; ,  \label{2thirtytwoa} \\
 {d^2 Y \over dt^2} + 2 n' {dX \over dt} &=& -{G_{12} m_0 \over r^3}
 Y \; .  \label{2thirtytwob}
\end{eqnarray}
\end{mathletters}
They admit (in the general case of a time-independent potential $F$)
the Jacobi energy integral
\begin{equation}
 {1\over 2}\left({\dot X}^2+{\dot Y}^2\right) - F(X,Y) = C = {\rm
 const.} \label{2thirtythree}
\end{equation}

\subsection{ Iterative solution of Hill's equations}
In spite of the apparent simplicity of Eqs.\ (\ref{2thirtytwo}) and
of the existence of the first integral (\ref{2thirtythree}), the
corresponding dynamics contains all the richness and complexity of
the three-body problem. Hill's idea was first to find an exact
{\it periodic} solution of Eqs.\ (\ref{2thirtytwo}) (with ${\bar
\delta}_{12} = 0$). The existence (in a mathematical sense) of
Hill's periodic solution, and the convergence of the power series
(in the parameter $m\equiv n'/(n-n')$) giving its explicit
form, have been proven by Wintner \cite{W26} (see \cite{Hagihara}
for more references). The existence of such a one-parameter family
of coplanar, periodic solutions, and the convergence of the
associated perturbation series in $m$, are stable under the
addition of the full series of higher-order tidal terms \cite{SM71}.
We are therefore confident that it will still exist when one adds
the ``dipole tidal'' term $F_1$, if ${\bar \delta}_{12}$ is small
enough.

To construct explicitly the perturbation series in $m$
giving the periodic solutions of Eqs.\ (\ref{2thirtytwo}) it
is convenient to rewrite them in terms of new variables. Following
a standard notation \cite{BC} (except that we do not introduce a
separate letter for the complex conjugate of $u$) we define
\begin{mathletters}
 \label{2thirtyfour}
\begin{eqnarray}
 u &=& X+ i Y \; , \label{2thirtyfoura} \\
 {\bar u} &=& X- i Y \; , \label{2thirtyfourb} \\
 \tau &=& (n-n')t + \tau_0 \; , \label{2thirtyfourc} \\
 \zeta &=& e^{i \tau} \; , \label{2thirtyfourd} \\
 D &=& {1 \over i} {d \over d\tau} = \zeta {d \over d\zeta}
 \; . \label{2thirtyfoure}
\end{eqnarray}
\end{mathletters}
Here, $n$ denotes the mean sidereal orbital velocity of the periodic
solution one is looking for (in other words the rotating frame
quantities $X,Y,u,{\bar u}$ are supposed to be periodic functions of
$\tau$ with period $2\pi$). The parameter $m =n'/(n-n')$ is
the remaining free parameter of the problem. It takes positive values
for prograde orbits (going in the same sense as the Sun\footnote{We
shall not consider the prograde orbits with $0<n<n'$ which are highly
unstable; see below.}: $0<n'<n$), and negative values
for retrograde orbits $n<0$. With this notation the general
equations of motion (\ref{2thirty}) read
\begin{equation}
 D^2u + 2m Du = - 2{\partial {\hat F} \over \partial{\bar u}}
 \label{2thirtyfive}
\end{equation}
(and its complex conjugate: $u \rightarrow {\bar u}$, $D \rightarrow
{\bar D} = -D$) where ${\hat F} \equiv (n-n')^{-2}F = m^2
n'^{-2} F$. For small values of $m$ (i.e. orbits of small
radius around the Earth) the periodic solution of Eq.\
(\ref{2thirtyfive}) is of the form $u \simeq ({\rm const.})\zeta$,
${\bar u}\simeq ({\rm const.}){\bar \zeta}=({\rm const.})\zeta^{-1}$.
It is convenient to replace $u$ by a variable $w$ which
tends to zero with $m$. Following Liapunov \cite{L96} and
Brumberg and Ivanova \cite{BI85}, one defines first a fiducial lunar
semi-major axis ${\tilde a}$ by writing
\begin{equation}
 {G_{12} m_0 \over (n-n')^2 {\tilde a}^3} = \kappa(m)
 \; , \label{2thirtysix}
\end{equation}
where
\begin{equation}
  \kappa(m) \equiv 1+2m +{3\over 2}m^2
 \; . \label{2thirtyseven}
\end{equation}
With this definition of ${\tilde a}$ one introduces the variable
$w$ by
\begin{mathletters}
 \label{2thirtyeight}
\begin{eqnarray}
 u &=& {\tilde a}\zeta (1+w) \; , \label{2thirtyeighta} \\
 {\bar u} &=& {\tilde a}\zeta^{-1} (1+{\bar w}) \; .
 \label{2thirtyeightb}
\end{eqnarray}
\end{mathletters}

The Lagrangian ${\cal L} = -2 {\tilde a}^{-2} (n-n')^{-2}
{\hat L}_{12}$ can be
written as
\begin{eqnarray}
 {\cal L}(w,{\bar w},Dw,D{\bar w}) &=& Dw D{\bar w} + 2(m+1)w D{\bar
 w} \label{2thirtynine} \\
 & & - G(w,{\bar w}) + ({\rm total\; derivative}) \nonumber \; ,
\end{eqnarray}
where $G \equiv 2{\tilde a}^{-2}{\hat F} + (1+2m)(1+w)(1+{\bar
w})$, and the associated equations of motion read
\begin{equation}
 D^2w + 2(m+1)Dw = - {\partial G \over \partial {\bar w}} \; .
 \label{2fourty}
\end{equation}

The explicit form of the potential $G(w,{\bar w})$ in our problem
(i.e. when $(n-n')^2 {\hat F} \equiv F = F_0+F_1+F_2$) reads
\widetext
\begin{eqnarray}
 G(w,{\bar w}) &=&  \kappa\left(m\right) \left[2\left(
  1+w\right)^{-1/2}\left(1+{\bar w}\right)^{-1/2}+\left(1+w\right)
  \left(1+{\bar w}\right)\right] + {\hat \lambda} \left[\zeta
  \left(1+w\right)+\zeta^{-1}\left(1+{\bar w}\right)\right]
  \nonumber \\
 & & + {3\over 4}m^2 \left[\zeta^2 \left(1+w\right)^2+
  \zeta^{-2}\left(1+{\bar w}\right)^2\right] \; ,
 \label{2fourtyone}
\end{eqnarray}
where we have introduced, instead of ${\bar \delta}_{12}$, the small
dimensionless parameter
\begin{equation}
 {\hat \lambda} \equiv m^2 {\bar \delta}_{12} {a' \over
 {\tilde a}}\; . \label{2fourtytwo}
\end{equation}
The corresponding equations of motion read
\begin{equation}
 D^2w + 2\left(m+1\right)Dw +  \kappa\left[1 + w -
 \left(1+w\right)^{-1/2} \left(1+{\bar w}\right)^{-3/2}\right]
 + {\hat \lambda} \zeta^{-1} + {3\over 2}m^2\zeta^{-2}
 \left(1+{\bar w}\right) = 0 \; . \label{2fourtythree}
\end{equation}
A last transformation \cite{L96}, \cite{BI85} consists of separating
off the square bracket multiplied by $\kappa$ in
(\ref{2fourtythree}) its nonlinear piece, namely
\begin{equation}
 Q(w,{\bar w}) \equiv \left(1+w\right)^{-1/2} \left(1+{\bar w}
 \right)^{-3/2} -1 + {1\over 2}w + {3\over 2}{\bar w}
 = {3\over 8} w^2 + {15\over 8} {\bar w}^2 + {3\over 4} w {\bar w}
  + {\cal O}\left(w^3,w^2{\bar w},w{\bar w}^2,{\bar w}^3\right)\; .
 \label{2fourtyfour}
\end{equation}
We can now define a {\it linear} operator acting on $(w,{\bar w})$,
\begin{equation}
 L(w,{\bar w}) \equiv D^2w +2\left(m+1\right) Dw + {3\over 2}
  \kappa\left(m\right)\left(w+{\bar w}\right) \; ,
 \label{2fourtyfive}
\end{equation}
and an effective source term (containing source terms and
non-linearities)
\begin{equation}
 W(w,{\bar w}) \equiv -{\hat \lambda}\zeta^{-1} - {3\over 2}
 m^2\zeta^{-2}\left(1+{\bar w}\right) + \kappa\left(
 m\right)Q\left(w,{\bar w}\right)\; . \label{2fourtysix}
\end{equation}
In terms of these definitions the equations of motion read
\begin{equation}
 L(w,{\bar w}) =  W(w,{\bar w})\; , \label{2fourtyseven}
\end{equation}
and its complex conjugate equation. This is the form used by Brumberg
and Ivanova \cite{BI85} in their study of relativistic effects in the
lunar motion.

Note that the source of all equivalence-principle violation
effects is the contribution $-{\hat \lambda}\zeta^{-1}$ on the
right-hand side of Eq.\ (\ref{2fourtysix}), with ${\hat \lambda}$
defined by Eq.\ (\ref{2fourtytwo}). Even when restricting oneself
(as we shall) to the effects linear in ${\hat \lambda}$, the
corresponding contributions in the solution are quite complicated
because of the interplay with the quadrupole tidal effects (from
$F_2$), i.e., mathematically, because of the mixed term $-\case3/2
m^2 \zeta^{-2} {\bar w}$ and the nonlinear term $\propto
Q(w,{\bar w})$ in $W(w,{\bar w})$.

When putting back the octupole tidal effects ($F_3$, Eq.\
(\ref{2twentynined})) they add to the effective source term
(\ref{2fourtysix})) the contribution
\begin{equation}
 W_3(w,{\bar w}) = -{3\over 8} {\hat \pi} \Bigl[\zeta(1+w)^2
 + 2\zeta^{-1}(1+w)(1+{\bar w}) + 5\zeta^{-3}(1+{\bar w})^2
 \Bigr]\; , \label{2fourtyeight}
\end{equation}
where
\begin{equation}
 {\hat \pi} \equiv m^2 \left(X_2-X_1\right) {{\tilde a} \over
 a'} = m^2 {m_2-m_1 \over m_2+m_1} {{\tilde a} \over a'} \; .
 \label{2fourtynine}
\end{equation}

The equation (\ref{2fourtyseven}) can be solved by iteration: first,
one keeps only the linear source terms which exist when $w=0$,
namely $W^{(1)}(\zeta,\zeta^{-1}) = W(0,0) = -{\hat \lambda} \zeta^{-1} -
\case3/2 m^2\zeta^{-2}$ [with the addition of $W_3(0,0) =
-\case3/8 {\hat \pi}(\zeta + 2\zeta^{-1} + 5\zeta^{-3})$ when
including parallactic terms]. Second, one solves the linear equations
$L(w,{\bar w}) = W^{(1)}(\zeta,\zeta^{-1})$ to get the corresponding
first-order solution: $w^{(1)} = {\hat \lambda}w^{(1)}_\lambda +
m^2 w^{(1)}_{m^2} + {\hat \pi}w^{(1)}_\pi$, which is
valid up to terms of higher order in the formal expansion parameters
${\hat \lambda}$, $m^2$ and ${\hat \pi}$. The next step is
to insert the first-order solution $w^{(1)}$ in the full source
term $W(w,{\hat w})$ and to collect the second-order source term
$W^{(2)}(\zeta,\zeta^{-1})$ of formal order $({\hat \lambda}+
m^2 + {\hat \pi})^2$. The corresponding second-order solution
$w^{(2)}$ is obtained by solving the linear equation $L(w^{(2)},
{\bar w}^{(2)}) = W^{(2)}(\zeta,\zeta^{-1})$, etc.. At each stage
of the iteration, one deals with a source term which is a linear
combination (with real coefficients\footnote{The reality of all
coefficients $W_k$ in Eq.\ (\ref{2fifty}) and $w_k$ in Eq.\
(\ref{2fiftyone}) is easily proven by induction, given the reality
of the coefficients in the exact definition of $W(w,{\bar w})$
and in the iterative solution Eq.\ (\ref{2fiftytwo}).}) of a finite
number of integer powers of $\zeta$ and $\zeta^{-1}$, say
\begin{equation}
 W_\star = W_0 + W_1\zeta + W_{-1}\zeta^{-1} + W_2\zeta^2 + W_{-2}
 \zeta^{-2} + \cdots + W_k\zeta^k + W_{-k}\zeta^{-k} \; .
 \label{2fifty}
\end{equation}
It is easy to check that there is a unique solution of the linear
system $L(w,{\bar w}) = W_\star$, ${\bar L(w,{\bar w})} = {\bar
W_\star}$ and that it is given by
\begin{equation}
 w_\star = w_0 + w_1\zeta + w_{-1}\zeta^{-1} + w_2\zeta^2 + w_{-2}
 \zeta^{-2} + \cdots + w_k\zeta^k + w_{-k}\zeta^{-k} \; ,
 \label{2fiftyone}
\end{equation}
with \cite{BI85}
\begin{mathletters}
 \label{2fiftytwo}
\begin{eqnarray}
 w_0 &=& {1\over 3\kappa(m)} W_0 \; ,\label{2fiftytwoa}
 \\
 w_k &=& {1\over \Delta_k} \left[\left(k^2 - 2\left(m+1\right)k +
 {3\over 2} \kappa\right) W_k - {3\over 2} \kappa W_{-k}
 \right]\; , \label{2fiftytwob} \\
 w_{-k} &=& {1\over \Delta_k} \left[\left(k^2 + 2\left(m+1\right)k
 +{3\over 2} \kappa\right) W_{-k} -{3\over 2} \kappa W_k
 \right]\; , \label{2fiftytwoc}
\end{eqnarray}
\end{mathletters}
Here $\Delta_k (=\Delta_{-k})$ denotes the determinant of the $2
\times 2$ system of equations satisfied by $w_k$ and $w_{-k}$ (when
$k\neq 0$). Its value is
\begin{equation}
 \Delta_k = k^2\left[k^2 +3 \kappa -4\left(m+1\right)^2
 \right] = k^2\left[k^2 -1 -2m+{1\over 2}m^2\right]\; .
 \label{2fiftythree}
\end{equation}
\narrowtext
\noindent This determinant never vanishes $(k\neq 0)$, but it takes a
small value of order $m$ when $k=\pm 1$. This small divisor is
one of the origins of the peculiar amplification which affects both
equivalence-principle-violation effects ($W^{(1)}(F_1)=-{\hat
\lambda}\zeta^{-1}$) and a part of the octupolar-tide effects
($W^{(1)}(F_3)=-\case3/8 {\hat \pi}(\zeta+2\zeta^{-1}+\cdots)$). A
consequence of the small divisor $\Delta_1 = -2m+\case1/2
m^2$ is that, when collecting from the iterative solution the
contributions proportional to ${\hat \lambda}\zeta^{\pm 1}$ and
${\hat \pi}\zeta^{\pm 1}$ they are found to proceed according to the
powers of $m^2/\Delta_1 ={\cal O}(m)$ instead of the
powers of $m^2$ as formally expected from the structure
(\ref{2fourtysix})\footnote{Following Poincar\'e \cite{Poin} (see also
Ref.~\cite{BC}), one can clarify the iterative process by giving a
new name to the parameter $m$ appearing in the second term of
$W$, Eq.\ (\ref{2fourtysix}), leaving unchanged the other
occurrences of $m$ in $L$ and $\kappa(m)$.}.

The first steps of the iteration can be done by hand. From Eqs.\
(\ref{2fiftytwo}) the linearized solution (without parallactic terms)
has the form
\begin{equation}
 w^{(1)} = w^{(1)}_1\zeta + w^{(1)}_{-1}\zeta^{-1}
  + w^{(1)}_2\zeta^2 + w^{(1)}_{-2}\zeta^{-2} \; ,
 \label{2fiftyfour}
\end{equation}
with
\begin{mathletters}
 \label{2fiftyfive}
\begin{eqnarray}
 w^{(1)}_1 &=& +{3\over 2}{\kappa \over \Delta_1} {\hat
  \lambda}\; ,\label{2fiftyfivea} \\
 w^{(1)}_{-1} &=& -{1+2(m+1)+\case3/2  \kappa \over
  \Delta_1} {\hat \lambda}\; ,\label{2fiftyfiveb} \\
 w^{(1)}_2 &=& +{9\over 4}{\kappa \over \Delta_2} m^2
  \; ,\label{2fiftyfivec} \\
 w^{(1)}_{-2} &=& -{3\over 2}{4+4(m+1)+\case3/2 \kappa
  \over \Delta_2} m^2\; .\label{2fiftyfived}
\end{eqnarray}
\end{mathletters}
Here, as defined above, $\kappa \equiv 1+2m+\case3/2
m^2$, $\Delta_1 = -2m + \case1/2 m^2$ and $\Delta_2
= 4 (3-2m +\case1/2 m^2)$. The insertion of the linearized
solution $w^{(1)}$, Eqs.\ (\ref{2fiftyfive}), into $W(w,{\bar w})$,
Eq.\ (\ref{2fourtysix}), generates a second-order source term with the
following structure: $W^{(2)} \sim {\hat \lambda}m^2(
\zeta^{\pm 1} + \zeta^{\pm 3})+ m^4 (\zeta^0+\zeta^{\pm 2} +
\zeta^{\pm 4}) + {\cal O}({\hat \lambda}^2)$. Let us focus on the
terms in the solution which are linear in ${\hat \lambda}$ and
contain the ``resonant'' frequencies $\zeta^{\pm 1}$. Their source
terms are found to be $W^{(2)}_1 ={99\over 64} m{\hat \lambda}\,
(1+{\cal O}
(m))$ and $W^{(2)}_{-1} = -{207\over 64}m{\hat \lambda}\,
(1+{\cal O}(m))$. From Eqs.\ (\ref{2fiftytwo}) the corresponding
solutions can be written as $w^{(2)}_1 = -{45\over 16}{\hat \lambda}
\,(1+{\cal O}(m))$, $w^{(2)}_{-1} = -3 w_1^{(2)} \,(1+{\cal
O}(m))$. At this approximation we have, when expanding
$w^{(1)}_{\pm 1}$ in powers of $m$,
\begin{mathletters}
 \label{2fiftysix}
\begin{eqnarray}
 w^{(1)}_1 + w^{(2)}_1 &=& \left(1 + {15\over 4}m + {\cal O}
  \left(m^2\right)\right) w^{(1)}_1 \nonumber \\
 &=& -{3\over 4} {{\hat \lambda} \over m}\left(1 + 6m
  + {\cal O}\left(m^2\right)\right) \; ,\label{2fiftysixa} \\
 w^{(1)}_{-1} + w^{(2)}_{-1} &=& \left(1 + {15\over 4}m +
  {\cal O}\left(m^2\right)\right) w^{(1)}_{-1} \nonumber \\
 &=& {9\over 4} {{\hat \lambda} \over m}\left(1 + {46 \over 9}
 m+{\cal O}\left(m^2\right)\right) \; .\label{2fiftysixb}
\end{eqnarray}
\end{mathletters}

\subsection{ Radial and angular perturbations due to a violation of
 the equivalence principle}
Let us relate the results (\ref{2fiftysix}) to the radial and
angular perturbations of the lunar motion associated with the
parameter ${\hat \lambda} \propto {\bar \delta}_{12}$. The radius
vector $r=(X^2+Y^2)^{1/2}$ and longitude $\theta$ of the Moon (with
respect to the rotating vector ${\bf e}_X = {\bf N}$, i.e. with
respect to the Sun) are such that $u=r e^{i\theta}={\tilde a}
\zeta(1+w)$, where we recall that $\zeta=e^{i\tau}$. Hence
\begin{mathletters}
 \label{2fiftyseven}
\begin{eqnarray}
 r^2 & = & u{\bar u} = {\tilde a}^2 (1+w)(1+{\bar w})\; ,
 \label{2fiftysevena} \\
 e^{2i\theta} & = & {u \over {\bar u}} = e^{2i\tau}
 {1+w \over 1+{\bar w}} \; . \label{2fiftysevenb}
\end{eqnarray}
\end{mathletters}

Working linearly in ${\hat \lambda}$ we get the following radial and
longitudinal equivalence-principle-violation perturbations
\begin{mathletters}
 \label{2fiftyeight}
\begin{eqnarray}
 {\delta_\lambda r \over {\tilde a}} & = & \Re e\,\left[\left(
  {1+{\bar w} \over 1+w}\right)^{1/2} \delta_\lambda w\right] \; ,
 \label{2fiftyeighta} \\
 \delta_\lambda \theta &=& \Im m\,\left[{\delta_\lambda w\over 1+w}
 \right]\; .  \label{2fiftyeightb}
\end{eqnarray}
\end{mathletters}

At the approximation (\ref{2fiftysix}) we can write $\delta_\lambda
r/{\tilde a} \simeq (w_1+w_{-1})\cos\tau + (w_3+w_{-3})\cos 3\tau$,
and $\delta_\lambda \theta \simeq (w_1-w_{-1})\sin\tau + (w_3-w_{-3})
\sin 3\tau$. In the approximation, the observable synodic effects are
entirely described by $w_1 \pm w_{-1}$. However, in higher
approximations, $w_{\pm3}, w_{\pm5}$, etc$\ldots$ feed down to the
synodic effects in $r$ and $\theta$. [Let us note in passing that,
when averaging over time, the mean shift of the cartesian
components $u=X+iY={\tilde a} \zeta (1+w)$ is given, to all orders,
by $w_{-1}$ alone: $\langle X \rangle = {\tilde a} w_{-1}$,
$\langle Y \rangle =0$.]

Focussing on the contributions at the synodic frequency
$n-n'$, we get at this stage (in agreement with Appendix A)
\widetext
\begin{mathletters}
 \label{2fiftynine}
\begin{eqnarray}
 \left({\delta_\lambda r \over {\tilde a}}\right)_{\rm synodic} & =
  & {3 \over 2} {{\hat \lambda} \over m}\left[1 + {14 \over 3}
  m+ {\cal O}\left(m^2\right)\right] \cos\tau\; ,
 \label{2fiftyninea} \\
 \left(\delta_\lambda \theta\right)_{\rm synodic} & = &
  -3 {{\hat \lambda} \over m}\left[1 + {16 \over 3}
  m+ {\cal O}\left(m^2\right)\right] \sin \tau\; .
 \label{2fiftynineb}
\end{eqnarray}
\end{mathletters}
\narrowtext

A straightforward, though slightly more involved, calculation allowed
us to compute by hand the ${\cal O}(m^2)$ contributions to the
square brackets on the right-hand sides of Eqs.\ (\ref{2fiftynine}).
In particular, we found that the square bracket in the range
perturbation, Eq.\ (\ref{2fiftyninea}), reads $1+{14\over 3} m+
{1061\over 48}m^2 + {\cal O}(m^3)$. In view of the large
coefficients appearing in this expansion, which create large
corrections to the lowest-order effect (for the Moon, ${14\over 3}
m=0.3773$ and ${1061\over 48}m^2=0.1445$), we have
decided to take
the bull by the horns and to solve iteratively the equations of
motion (\ref{2fourtyseven}) to a very high order by using the
dedicated computer manipulation programme MINIMS written by M. Moons
from the University of Namur (Belgium) (see \cite{Moons}). Some
details on the application of this programme to our problem are
given in Appendix~B. Let us quote here the form of the results.
Replacing ${\hat \lambda}$ by its definition (\ref{2fourtytwo}), we
see that ${\tilde a}$ drops out when writing the range perturbation
$\delta_\lambda r$. Finally, we can write
\begin{mathletters}
 \label{2sixty}
\begin{eqnarray}
 \left(\delta_\lambda r\right)_{\rm synodic} & = & C\left(m
 \right) {\bar \delta}_{12} a' \cos\tau\; , \label{2sixtya} \\
 \left(\delta_\lambda \theta\right)_{\rm synodic} & = &
 -C'\left(m\right) {\bar \delta}_{12} {a' \over {\tilde a}}
 \sin \tau\; , \label{2sixtyb}
\end{eqnarray}
\end{mathletters}
where
\begin{mathletters}
 \label{2sixtyone}
\begin{eqnarray}
 C\left(m\right)&=& {3\over 2}m\left(1 + \sum_{k\geq 1}
  c_k m^k\right) \equiv {3\over 2}m\, S\left(m
 \right) \; , \label{2sixtyonea} \\
 C'\left(m\right)&=& 3 m\left(1 + \sum_{k\geq 1}
  c'_k m^k\right) \equiv 3 m\, S'\left(m\right)\; .
 \label{2sixtyoneb}
\end{eqnarray}
\end{mathletters}

The beginning of the power series $S(m)$ entering the synodic
range perturbation is
\begin{eqnarray}
 S(m)& =&1+ {14\over 3}m+ {1061\over 48}m^2
  + {2665\over 24}m^3 +{145683\over 256}m^4
 \nonumber \\
 & & + {6729119\over 2304}m^5 + {1656286531\over 110592}
 m^6 + \cdots \; . \label{2sixtytwo}
\end{eqnarray}
The coefficients of the series $S(m)$ and $S'(m)$ are
given in Appendix~B up to the power $m^{17}$ included. They are found
to grow fast. The ratio between two successive coefficients $c_k/
c_{k-1}$ or $c'_k/c'_{k-1}$ is found, numerically, to converge
rapidly to the value $5.1254717\ldots$, thereby mimicking a
geometric series in $m/m_{cr}$ with $m_{cr}=
(5.1254717\ldots)^{-1} \simeq 0.195103996\ldots$. In the case of the
Moon, with $m =0.0808489375\ldots$ \cite{GS86}, this means
that the series $S(m)$ and $S'(m)$ converge rather
slowly, as geometric series of ratio $m/m_{cr} \simeq
0.4144$. The truncation to order $m^{17}$ is just enough
to estimate the values of the series to the $10^{-5}$ accuracy.
We shall discuss below a method which allows us to improve this
precision. We find, for the Moon,
\begin{equation}
 S=1.62201\ldots \;\; ,
S'=1.72348\ldots \;,
\end{equation}
 so that the
full coefficients appearing in the synodic effects (\ref{2sixty})
are, respectively, $C=0.196707\ldots$, $C'=0.418025\ldots$.
Finally, using the recommended value of the semi-major axis of
the Earth orbit, $a' = a_{Earth}A = 1.495980221\times 10^{13}\,
{\rm cm}$ \cite{CT95} (where $A$ denotes the astronomical unit),
the amplitude of
the range oscillation of the Moon due to an
equivalence principle violation is numerically found to be
\begin{equation}
 Ca' {\bar \delta}_{12} = 2.9427\times 10^{12} {\bar \delta}_{12}
 \;\; {\rm cm}\; . \label{2sixtythree}
\end{equation}
In the case where one assumes the absence of violation of the weak
equivalence principle, i.e. ${\hat \delta}_A \equiv 0$ in Eq.\
(\ref{2ten}), the result (\ref{2sixtythree}) gives
\begin{equation}
( \delta_\lambda r)_{\rm synodic} = 13.10 \eta \cos\tau\;
{\rm meters}\; ,
\end{equation}
 if we use Eq.\ (\ref{2five}) \cite{WND95} as nominal value for the
difference of gravitational binding energies. Our final result is
approximately $60$ \% larger than the lowest-order estimate first
derived by Nordtvedt in 1968 \cite{N68c} and recalled in Eqs.\
(\ref{1four}), (\ref{1five}) and (\ref{1eight}) above. On the other
hand, it confirms the recent finding of Nordtvedt \cite{N95} that the
interaction with the orbit's tidal deformation significantly amplifies
the synodic range oscillation and substantiates it by providing
explicit and accurate expressions for the amplitude of the synodic
effect.

\section{ Physical discussion}

\subsection{ Resonances and instability}
We have seen in the previous section that the series in powers of
$m$ giving the amplitudes of synodic perturbations
(\ref{2sixty}) appear to be close to geometric series in $m/
m_{cr}$ with $m_{cr} \simeq 0.195104$. This suggests
the existence of pole singularities $\propto (m_{cr}-
m)^{-1}$ at $m=m_{cr} \simeq 0.195104$ in those
series. Nordtvedt \cite{N95} suggested also the presence of such a pole
singularity at $m\simeq 0.2$ (i.e. for a sidereal period of
about $2$ months) on the basis that for such a high orbit the driving
frequency ($n'$ in a non-rotating frame) might become equal to the
rate of perigee advance $(d\varpi/dt)$. We have substantiated and
generalized this suggestion, as well as obtained by several
independent approaches a much more precise value for $m_{cr}$,
namely,
\begin{equation}
 m_{cr} = 0.1951039966\ldots \label{3one}
\end{equation}
by making use, notably, of the work of H\'enon \cite{Henon} on
the three-body
problem. To relieve the tedium the details of our arguments are
relegated to Appendix~C. Let us summarize our approach and our
results.

Our approach consists of putting together the (numerical) results of
H\'enon \cite{Henon} on the stability of the periodic orbits in
Hill's problem, with some knowledge of the general structure of
Hamiltonian perturbations, and a more specific use of the analytical
structure of the solutions of Hill's variational equations in presence
of ``forcing'' terms, such as the ones coming from the potentials $F_1$,
Eq.\ (\ref{2twentynineb}), and $F_3$, Eq.\ (\ref{2twentynined}),
which are neglected in Hill's main problem. Our conclusions are that
when $m$ increases up to $m_{cr}$, Eq.\
(\ref{3one}), there is a confluence of correlated singularities: on
the one hand, as found by H\'enon, the free perturbations of Hill's
orbit (those not driven by any additional forces) loose their
stability, and on the other hand, all perturbations driven by
perturbing potentials of any odd frequency in the rotating frame
(i.e. containing terms $\propto \exp[\pm(2k+1)i\tau]$) develop pole
singularities $\propto (m_{cr}-m)^{-1}$. As indicated
by Nordtvedt the value $m=m_{cr}$ does correspond to
a simple $1:1$ commensurability $d\varpi/dt = n'$ or ${\rm c}
\equiv dl/d\tau =1$ in the rotating frame (where $l=nt+\epsilon -
\varpi$ is the mean anomaly). Note, however, that, contrary to what
happens in the familiar case of a harmonic oscillator, the basic
frequency of the driving force does not need to have a $1:1$
resonance with the natural frequency of the orbit (perigee
precession) to generate poles $\propto (m_{cr}-m)^{-1}$;
the odd commensurabilities $3:1$, $5:1$, etc. generate similar
poles. Therefore, both the (hypothetical)
equivalence-principle-violation effects (potential $F_1$), and many
of the (really
existing) parallactic effects (potentials $F_3$, $F_5$, $\ldots$)
will have pole singularities $\propto (m_{cr}-m)^{-1}$.
Moreover, these poles are present not only in the synodic terms
(that we concentrate upon here) but in the terms at frequencies
$3(n-n')$, $5(n-n')$, $\ldots$. The consequences of this situation
are explored in the following subsections.

\subsection{ Pad\'e approximant of equivalence-principle-violation
 effects}
The analysis of Appendix~C shows that the amplitudes of the synodic
perturbations (\ref{2sixty}) considered as functions of $m$
have a simple pole (but no branch point) on the positive real axis
at $m=m_{cr}$, Eq.\ (\ref{3one}), and have no
singularities on the negative real axis down to $m=-1$
(because of the stability of the retrograde orbits \cite{Henon})
\footnote{The value $m=-1$ corresponds to very wide retrograde
orbits $0<-n\ll n'$.}.
This simple analytical behaviour suggests that the numerical validity
of the power series representation (\ref{2sixtyone}) can be
efficiently extended by using Pad\'e approximants, i.e. by
rewriting the power series $S(m)$, $S'(m)$ truncated
at order $m^{17}$ as quotients $N(m)/D(m)$,
$N'(m)/D'(m)$ of two power series truncated at order
$m^8$. The explicit coefficients of the Pad\'e approximants
\footnote{It is to be noted that, thanks to the nearly
geometric progression of the coefficients in many of the power
series of Appendix B, a simpler (though less general)
alternative to Pad\'e approximants would be simply to
factorize $(1 - m/m_{cr})^{-1}$.}
are given in Appendix~B. We have done several checks of the
conjecture that these Pad\'e approximants provide a numerically
accurate representation of the exact solution $S(m)$ on the
entire interval $(-1,m_{cr})$. First, the real zeros of
smallest absolute value of the denominators $D(m)$ and
$D'(m)$ are respectively found to be $0.19510399668\ldots$
and $0.19510399660\ldots$ in excellent agreement with H\'enon's
value (\ref{3one}). Second, we found that the Pad\'e approximants
truncated to order $m^7$ numerically agree all over the
interval $(-1,m_{cr})$ with those at order $m^8$ within
better than $1$ \%. [Actually, the difference is much smaller
than $10^{-3}$ except very near $m=-1$.] We plot in Fig.~1
the Pad\'e approximant of the coefficient $C(m)$ of the
radial synodic effect,
\begin{equation}
 C_{\rm Pade}(m) = {3\over 2} m {N_8(m) \over
  D_8(m)}\; , \label{3two}
\end{equation}
over the interval $(-1,m_{cr})$. Let us note the two numerical
values (using $m_{\rm Moon} = 0.0808489375\ldots$; \cite{GS86})
\begin{mathletters}
 \label{3three}
\begin{eqnarray}
 C_{\rm Pade}(m_{\rm Moon})&=& 0.196707 \; , \label{3threea}\\
 C_{\rm Pade}(-1) &=& -0.267706 \; . \label{3threeb}\\
\end{eqnarray}
\end{mathletters}

In Fig.1 the lunar value (3.3a) is indicated by a dot. The
difference between the linearized approximation ${3\over 2}
m$ (dashed line in Fig.1) and the exact value of $C(m)$ (solid line)
illustrates the importance of nonlinear effects in the radial
synodic perturbation.

\begin{figure}
\caption{Coefficient $C(m)$ of the synodic range oscillation
 (defined in Eq.~(2.60a)) as a function of $m=n'/(n-n')$.
 The solid line represents the Pad\'e approximant of $C(m)$,
 while the dashed line represents the linearized approximation
 $\case3/2 m$. The dot indicates the actual lunar value.}
\label{fig1}
\end{figure}

\subsection{ Better orbital tests of the equivalence principle?}
Nordtvedt \cite{N95} has suggested to make use of the resonance at
$m=m_{cr}$ to improve the precision of equivalence
principle tests. The idea would be to put an artificial satellite
in an orbit close to the resonant orbit ($m=m_{cr}$).
{} From our numerical estimates, the resonant orbit has a sidereal
period $T_{cr}=m_{cr} (1+m_{cr})^{-1} T'$ (where $T'=
2\pi/n' = 1\; {\rm year}$), i.e. $T_{cr}=1.95903$ month. The
corresponding ``bare'' semi-major axis $a_0\equiv(Gm_0/n^2)^{1/3}$
is $1.68255\; a_0({\rm Moon})$. Though interesting, this suggestion
is fraught with difficulties. First, our results show that one must
be very careful to use a {\it sub-critical} orbit ($m<
m_{cr}$) as super-critical orbits are exponentially unstable
(real characteristic multiplier $>1$). Second, the fact that all
the parallactic perturbations (proportional to $m_3^{\rm grav}/a'^k$
with $k\geq 4$, i.e. to $m^2 ({\tilde a}/a')^{k-3}$) develop
also pole singularities at $m=m_{cr}$ probably implies
that the orbit becomes unstable slightly below the ideal Hill
value (\ref{3one})\footnote{Unpublished calculations of H\'enon
(private communication) for a small but non-zero mass ratio
$\mu=m_0/(m_0+m_3)=10^{-6}$ show that the topology of the loss
of stability of Hill's prograde orbits is different from the ideal
Hill case ($\mu=0$) and the same as for generic values of $\mu
\neq 0$. The difference takes place in a region of fractional size
$10^{-3}$ ($\sim \mu^{1/2}$?) which suggests that the actual
$m_{cr}$ is roughly $0.1$ \% smaller than the value
(\ref{3one}).}. Moreover, the blow up of the parallactic perturbations
amplify already large synodic effects which are known only with
finite accuracy. Indeed, for some years the precision of the Lunar
Laser Ranging experiment has been limited by the accuracy with
which one could theoretically compute and subtract the synodic
parallactic perturbations \cite{S76}. To investigate this point
we have included the octupole-tide perturbation $F_3$, Eq.\
(\ref{2twentynined}), i.e. we added the contribution $W_3$, Eqs.\
(\ref{2fourtyeight}), in our Hill-Brown iteration program. Our
explicit results are given in Appendix~B. The form of the radial
perturbation is
\widetext
\begin{mathletters}
 \label{3four}
\begin{eqnarray}
 \left(\delta_\pi r\right)_{\rm synodic} &=& {15\over 16} {m_2-m_1
 \over m_2+m_1} \left({m_1+m_2 \over m_1+m_2+m_3}\right)^{2/3} a'
 P\left(m\right) \cos\tau \; , \label{3foura} \\
 P\left(m\right) &=& m^{7/3} \left[\kappa\left(
 m\right)\right]^{-2/3} Q\left(m\right) \; ,
 \label{3fourb} \\
 Q\left(m\right) &=& 1 + {22\over 5}m+ {215\over 16}
 m^2+ {57599 \over 960}m^3 + {917401 \over 2880}m^4
 + {230247737\over 138240}m^5 + {14206254151\over 1658880}
 m^6 + \cdots\; . \label{3fourc}
\end{eqnarray}
\end{mathletters}
\narrowtext
The numerical value of the coefficient giving the scale of
$\delta_\pi r$ in Eq.\ (\ref{3foura}) is $\delta_\pi r\simeq 28716.38
P(m)\cos\tau$ kilometers. The Pad\'e approximant of the
series $Q(m)$ is given in Appendix~B. For rough orders of
magnitude estimates we can approximate (when $0<m<
m_{cr}$) $P(m)$ by $P(m) \simeq m^{7/3}\,
(1-m/m_{cr})^{-1}$. By comparison, the coefficient
entering the Nordtvedt effect (\ref{2sixtya}) can be roughly
approximated by $C(m) \simeq \case3/2 m\,(1-m/
m_{cr})^{-1}$. Let us define the {\it amplification factor}
of the Nordtvedt effect as the ratio $A(m)\equiv C(m)/
C(m_{\rm Moon})$ where $C(m)$ is the coefficient in Eq.\
(\ref{2sixtya}). The amplification factor in the synodic
parallactic oscillation (\ref{3foura}) will be $B(m)\equiv
P(m)/P(m_{\rm Moon}) \simeq (m/m_{cr})^{4/3}
A(m)$. For an artificial satellite ($m_1 \ll m_2$)
one expects from Eq.\ (\ref{3foura}) that the main uncertainty in
the theoretical value of $\delta_\pi r$ will come from the
Earth/Sun mass ratio: $m_2/m_3$. The current fractional
uncertainty on this
ratio is $\epsilon_8\, 10^{-8}$ with $\epsilon_8 \simeq 1$
\cite{Science}. The corresponding uncertainty in $\delta_\pi r$
is $0.073\epsilon_8\, B(m)$ centimeters. Therefore the use
of a higher orbit, amplifying the Nordtvedt effect by a factor
$A(m)$, will entail a correspondingly increased uncertainty
on the synodic parallactic radial oscillation:
\begin{equation}
 \delta_\pi r \simeq 0.073\epsilon_8 \left(m/m_{\rm Moon}
 \right)^{4/3} A\left(m\right) \cos \tau \; {\rm cm}\; .
\label{3five}
\end{equation}
The problem might cure itself by the fact that the ratio $m_2/m_3$
will enter also the $\cos 3\tau$ parallactic effects, which will be
also amplified. But things might get complicated because, as one
approaches the resonance, several frequencies become close to each
other and one needs long data span to resolve the various frequencies
and measure separately their Fourier coefficients. Moreover, the
real motion of an artificial satellite beyond the Moon's orbit will
be very complex because of the combined gravitational effects of the
Earth and the Moon. Finally, such a satellite would need to be
endowed with a very high performance drag free system to compete with
the Moon which is, naturally, drag free to high precision.

In view of the difficulties associated with near-resonant lunar-type
orbits it is worth thinking about other possibilities\footnote{Let
us note in passing that, because of tidal dissipation, the Moon itself
is, kindly, slowly receding toward higher orbits. However, even under
the overoptimistic assumption that the present rate of energy
dissipation continues to apply in the future, the increase in the
semi-major axis
of the Moon will be only $\simeq 23$ \% in $6$ billion years (which
is the expected lifetime of the Sun)\cite{Laskar}.}. Let us list
some possibilities: artificial satellites around outer planets would
be interesting in that the basic dimensionful scale factor in the
synodic effect (\ref{2sixtya}) is $a'$, the semi-major axis of the
considered planet around the Sun. That would give a factor $5$ for
Jupiter and a factor $10$ for Saturn. In either case one would need
far enough satellites (i.e. $m$ big enough) to have a
coefficient $C(m)$ at least comparable to the lunar value
(\ref{3three}). A second possibility is to use retrograde orbits
which are always stable (in the Hill approximation). However, Fig.~1
shows that they give, at best, a factor $C(-1)=-0.267706$. An
equivalence-principle mission consisting of a pair of artificial
satellites around an outer planet (one prograde, one retrograde),
with a laser link between the satellites, could improve by a
significant factor upon the LLR experiment. Besides an improved
scale factor $a'$, the advantage of being around an outer planet
is that the radiation pressure from the Sun is much
smaller, so that the requirements on the drag-free system are
much less stringent.

\subsection{ Theoretical significance of orbital tests of the
 universality of free fall}
As we mentioned above, the Lunar Laser Ranging experiment is
sensitive, through the synodic effect (\ref{2sixty}), to the
sum of two physically independent contributions
\begin{eqnarray}
 {\bar \delta}_{12} &=& {\hat \delta}_1 - {\hat \delta}_2 + \eta
 \left({E^{\rm grav}_1 \over m_1 c^2} - {E^{\rm grav}_2 \over m_2
 c^2}\right) \nonumber \\
 &=& {\hat \delta}_{12} + 4.45\,\eta\, 10^{-10}\; . \label{3six}
\end{eqnarray}
The first contribution, ${\hat \delta}_{12} \equiv {\hat \delta}_1
- {\hat \delta}_2$, is essentially equivalent to what Newton and
Laplace had in mind when they proposed orbital tests of the
universality of free fall: bodies of different internal compositions
could fall differently. The second contribution, proportional
to $\eta = 4{\bar \beta}-{\bar \gamma}$, was discovered by Nordtvedt
who was assuming that the most natural theoretical framework in
which to study possible macroscopic deviations from Einstein's
theory is the class of metrically-coupled theories of gravity (see
e.g. \cite{W81} for a review). Actually, from the perspective of modern
unified theories the class of metrically-coupled theories of gravity
seems quite ad hoc. For instance, string theory does suggest the
possibility that there exist long-range scalar fields contributing
to the interaction between macroscopic bodies and thereby modifying
the standard predictions of general relativity. However, all the
scalar fields present in string theory have composition-dependent
couplings for very basic reasons (for a discussion of
general theoretical alternatives to Einstein's theory and the types
of composition-dependent couplings they might exhibit see
\cite{LH95}).

Recently, a mechanism has been proposed by which some
of the scalar fields of string theory might survive in the
macroscopic world as very weakly coupled long range fields
\cite{DP94} (see also \cite{DN93}).
 In the model of Ref.~\cite{DP94} the surviving scalar
field(s) modify the observational consequences of general relativity
in several ways: (i) they violate the ``weak equivalence principle''
(${\hat \delta}_A \neq 0$) because of the composition-dependence
of their couplings to matter; (ii) they modify the post-Newtonian
(${\cal O}(1/c^2)$) effects in essentially the way which is
parametrized by the Eddington parameters\footnote{This comes from
a feature of their couplings which is deeply rooted into the
structure of QCD and the consequences it has for the mass of atoms;
see pages 550--553 of \cite{DP94}.} ${\bar \beta}$ and ${\bar
\gamma}$; and (iii) they induce a slow time variability of all
the coupling constants of Nature: $G$, $\alpha$, $\alpha_{weak}$,
$\ldots$ . The point we want to emphasize here, because we think it
is generic, is that all those modifications of general relativity are
related, because they derive ultimately from the couplings of the
same field. In particular, the first term on the right-hand side
of Eq.\ (\ref{3six}) is related to the second.

More precisely, in the model of Ref. \cite{DP94}, we have,
for an individual atom labelled by $A$, the link
\begin{equation}
\hat{\delta}_A \simeq -0.943 \times 10^{-5} \bar{\gamma} (E/M)_A
\label{3seven}
\end{equation}
where $E \equiv Z(Z-1)/(N+Z)^{1/3}$ is associated to the electrostatic
interaction energy of the nucleus of the atom, and where $M$ denotes
the mass of $A$ in atomic mass units.
 We believe that the structure of this link between $\hat{\delta}_A$
and $\bar{\gamma} (E/M)_A$\footnote{We neglect here contributions
proportional to the ratios (baryon number)/(mass) and
(neutron excess)/(mass) which tend to be subdominant, even for
moderate $Z$ differences \cite{DP94}.} is generic in string-derived
models, because it follows from a basic physical feature of the massless
scalar fields (``moduli'') present in string theory, namely that
they define the values of the gauge coupling constants. Even the
magnitude of the numerical coefficient shoud be somewhat generic.
Indeed, its analytical expression $-{1\over 2} a_3 \alpha
\lambda_{\alpha} / \lambda_{u_3}$ (in the notation of \cite{DP94})
shows that it is determined by basic physical facts or assumptions:
fractional smallness of electrostatic nuclear contributions (
$a_3 \alpha \simeq 0.770 \times 10^{-3}$), unification of gauge
coupling constants ($\lambda_{\alpha} \simeq 1$) and QCD confinement
($\lambda_{u_3} \simeq \ln (\Lambda_{\rm string} / ({\rm a.m.u.}))
\simeq 40.8$).

We have also the model-dependent link ${\bar \beta} \simeq -10.2
\kappa {\bar
\gamma}$, where the dimensionless theory parameter $\kappa$ is expected
to be of order unity. [$\kappa >0$ denotes here the curvature of a
coupling function around a minimum and should not be confused
with the notation $\kappa (m)$ used above.]
These links indicate that, in the Earth-Moon case, the
gravitational binding contribution to ${\bar \delta}_{12}$ is
numerically negligible compared to the composition-dependent term
${\hat \delta}_{12}$. Indeed, using Eq. (\ref{3seven}) and the
compositional difference between the Earth and the Moon (i.e. the
difference between an Earth iron core of mass $0.32 m_2$ and a
silica-dominated Moon \cite{Su94}), we find $\hat{\delta}_{12}
\simeq 0.32(\hat{\delta}_{\rm Si}-\hat{\delta}_{\rm Fe})\simeq 2.75
\times 10^{-6} \bar{\gamma}$, while the gravitational binding
energy contribution is $4.45 \eta \times 10^{-10} = -4.45 (40.8
\kappa +1) \times 10^{-10} \bar{\gamma}$.
{}From the point of view advocated here, the conclusion is that LLR
data give us a very precise test of the {\it weak} equivalence
principle. The loss of a Nordtvedt-type direct test of the
combination $\eta = 4{\bar \beta} - {\bar \gamma}$ is
compensated by the theoretically expected link
${\bar \delta}_{12} \simeq {\hat \delta}_{12} \simeq 2.75 \times
10^{-6} {\bar \gamma}$ which gives an extremely good limit on the
effective Eddington parameter of the considered scalar model.
More precisely, the observational
limit\footnote{This was obtained from partial derivatives of the
numerically integrated equations of motion, and therefore
independently of theoretical estimates of the value of the coefficient
$C(m)$ in Eq.\ (\ref{2sixtya}).} ${\bar \delta}_{12}= (-3.2
\pm 4.6) \times 10^{-13}$ recently derived from LLR data \cite{Science},
\cite{WND95} translates into the following observational constraint
\footnote{We do not take into account here the theoretical constraint
that $\bar{\gamma} <0$ in all scalar models.} on $\bar{\gamma}$:
\begin{equation}
\bar{\gamma} = (-1.2 \pm 1.7) \times 10^{-7} \; . \label{3eight}
\end{equation}
The recent laboratory tests of the weak equivalence principle give
comparable results. Using the experimental limit
$\hat{\delta}_{{\rm Be \, Cu}} = (-1.9 \pm 2.5) \times 10^{-12}$
\cite{Su94} and the theoretical formula (\ref{3seven}) (which yields
$\hat{\delta}_{{\rm Be \, Cu}} = 2.41 \times 10^{-5} \bar{\gamma}$)
we find
\begin{equation}
\bar{\gamma} = (-0.8 \pm 1.0) \times 10^{-7} \; . \label{3nine}
\end{equation}
Impressive as these limits
may seem, Ref.~\cite{DP94} gives a motivation for pushing
equivalence principle tests further because this reference estimates
that a natural range for the coupling parameter ${\bar \gamma}$
is $10^{-19} \lesssim -{\bar \gamma} \lesssim 10^{-10}$.
[Note, however, that if the theory parameter $\kappa$
is of order $1/40$ (which corresponds, in the notation of
\cite{DN93}, to $\kappa \sim 1$) larger values of $- \bar{\gamma}$,
of order $10^{-7}$, are expected, in agreement with \cite{DN93}.]
In this  connection,
let us mention that the LLR CERGA team plans to improve the precision
of the ranging down to the 2-3 millimeter level for normal points
[C. Veillet,
private communication]. Extracting ${\bar \delta}_{12}$ at this level
will
necessitate to improve the modelling of the solar radiation
pressure effects which are currently believed to contribute a
synodic range oscillations of approximately $0.3$ cm \cite{N95}.
If this can be done, the LLR experiment will reach the level
${\bar \delta}_{12} \sim 5\times 10^{-14}$ corresponding to the level
$\bar{\gamma} \sim 10^{-8}$.
 It seems that significant progress
in testing the equivalence principle will require space missions:
either a low Earth orbiting artificial satellite dedicated to testing the
weak equivalence principle, as the STEP (Satellite Test of the
Equivalence Principle) mission, or, possibly, some type of orbital
test such as the one suggested in Ref.~\cite{N95} and the ones suggested
above.

\acknowledgments

We are grateful to M. Moons for providing us with the algebraic
manipulator MINIMS. We thank V.I. Arnold, V.A. Brumberg, J. Chapront,
B. Chauvineau, M. H\'enon, J. Henrard, J. Laskar, F. Mignard,  D. Ruelle,
H.H. Rugh, D. Sullivan and
C. Veillet for informative discussions or communications.  D. V. worked
on this paper while staying at the OCA/CERGA, Grasse (France) and being
supported by an H. Poincar\'e fellowship. He is also grateful to IHES,
Bures sur Yvette (France) for its kind hospitality and partial support.

\appendix
\section{ Traditional lunar perturbation theory}

As a check on the lowest orders of the Hill-Brown calculations
presented in the text, we have also investigated the mixing between
equivalence-principle-violation effects and tidal effects by means
of the
traditional lunar perturbation theory of de Pont\'ecoulant
\cite{dP}, \cite{Brown}. The equations of motion corresponding to
the Lagrangian ${\hat L} = \case1/2 {\bf v}^2 + \mu/r + R$ read
\begin{equation}
 {d^2 {\bf r}\over dt^2}+\mu{{\bf r}\over r^3} = {\partial R\over
 \partial {\bf r}}\; . \label{aone}
\end{equation}
Here, ${\bf r}\equiv {\bf x}_{12}\equiv {\bf x}_1-{\bf x}_2$ is the
position vector of the Moon with respect to the Earth (in an
inertial, non-rotating, coordinate system), $\mu\equiv G_{12}(m_1+
m_2)$ and $R = R_1+R_2+R_3+\cdots$ is the total potential perturbing
the Keplerian motion of the Moon around the Earth. [This corresponds
to Eqs.~(\ref{2twentyone}), (\ref{2twentytwo}).] We consider the
coplanar problem for which it is enough to solve for the radius $r
\equiv |{\bf r}|$ and the longitude $v$ (polar angle). Decomposing
the acceleration into radial and longitudinal components leads to
\begin{mathletters}
 \label{atwo}
\begin{eqnarray}
 {\ddot r} - r {\dot v}^2 &=& -\mu r^{-2} + \partial R/\partial r
 \; , \label{atwoa} \\
 d(r^2 {\dot v})/dt &=& \partial R/\partial v \; . \label{atwob}
\end{eqnarray}
\end{mathletters}
de Pont\'ecoulant's method uses the longitudinal equation
(\ref{atwob}) but replaces the radial one (\ref{atwoa}) by the
``virial'' equation dealing with the second time derivative of
$r^2$. The basic equations are then written as
\begin{mathletters}
 \label{athree}
\begin{eqnarray}
 {1\over 2}{d^2 \over dt^2}\left(r^2\right) - {\mu \over r} +
 {\mu\over a_c} &=& P \; , \label{athreea} \\
 {\dot v} -{h_c\over r^2} &=&{1\over r^2}\int dt\, {\partial R\over
 \partial v} \; , \label{athreeb}
\end{eqnarray}
\end{mathletters}
where $a_c$ and $h_c$ are some integration constants and where the
transformed source term in the radial equation is
\begin{eqnarray}
 P &\equiv & r{\partial R\over \partial r} + 2\int dt \left({d
 \over dt}\right)_1 R \nonumber \\
 &=& r{\partial R\over \partial r} + 2R +2n'\int dt\,{\partial R\over
 \partial v} \; . \label{afour}
\end{eqnarray}
In the first form of $P$, $(d/dt)_1$ denotes a time derivative taking
into account only the variability due to the time-dependence of the
coordinates of the Moon: $(d/dt)_1={\dot r}\partial/\partial r+
{\dot v}\partial/\partial v$. The second form of $P$ is obtained
by taking into account the time-dependence of $R$ upon the Sun's
coordinates, and assumes that the Sun moves on a circular orbit
(${\dot r}'=0$, ${\dot v}' = n'$). It is very useful to notice that
if $R=\sum_p R_p$ where each contribution $R_p(r,v-v')$ has a
radial dependence $\propto r^p$, then $P=\sum_p P_p$ with
\begin{equation}
 P_p = \left(p+2\right)R_p + 2n'\int dt\, {\partial R_p\over
 \partial v} \; . \label{afive}
\end{equation}
Note that the use of the suffix $p$ is consistent with the notation
$R_1, R_2, R_3$ of Eqs.~(\ref{2twentytwo}).

First-order perturbation theory is very easy. Let us consider a
general term $R^{(q)}_p=Ar^p\cos q(v-v')$, perturbing the zeroth-order
(circular) solution $r=a$, $v_0 = nt+\varepsilon$ (with zeroth-order
integration constants $a_c=a$, $h_c=na^2$, and the link $n^2a^3=
\mu$). Inserting the perturbed solution $r=a+\delta r$, $v=v_0+\delta
v$ into Eqs.~(\ref{athree}) (with perturbed integration constants
$a_c=a+\delta a$, $h_c = h+\delta h$) yields
\widetext
\begin{mathletters}
 \label{asix}
\begin{eqnarray}
 {d^2\over dt^2}\left(\delta r\right) + n^2\delta r &=& {\tilde p}
 A a^{p-1} \cos\left[q\left(n-n'\right)\left(t-t_0\right)\right] +
 n^2 \delta a \; , \label{asixa} \\
 {d \over dt}\left(\delta v\right) + 2 {n\over a}\delta r &=&
 {1\over (n-n')}A a^{p-2}\cos\left[q\left(n-n'\right)\left(t-t_0
 \right)\right] + a^{-2}\delta h \; , \label{asixb}
\end{eqnarray}
\end{mathletters}
where
\begin{equation}
 {\tilde p}\equiv p+2 +2{n'\over n-n'} = p+2+2m \; .
 \label{aseven}
\end{equation}
Here, as in the text, we shall use $m\equiv n'/(n-n')$ as the
small expansion parameter of perturbation theory. The solution of
Eqs.~(\ref{asix}) is (with $\tau\equiv v_0-v_0' = (n-n')(t-t_0)$)
\begin{mathletters}
 \label{aeight}
\begin{eqnarray}
 \delta r &=& {{\tilde p}\over n^2-q^2(n-n')^2}Aa^{p-1}\cos q\tau
 + \delta a \; , \label{aeighta} \\
 \delta v &=& -{1\over q(n-n')}\left[{2{\tilde p}n\over n^2-q^2(n-
 n')^2}-{1\over n-n'}\right] Aa^{p-2}\sin q\tau +(\delta n)t +
 \delta \epsilon \; , \label{aeightb}
\end{eqnarray}
\end{mathletters}
\noindent where $\delta n = -2na^{-1}\delta a + a^{-2}\delta h$.
Eqs.~(\ref{asix})--(\ref{aeight}) have been written assuming $q\neq
0$. They take a different form when $q=0$. It is traditional
\cite{Brown} to keep $n$ fixed throughout the approximation
process (and therefore numerically equal to the observed mean
motion), and to define $a$ by $n^2 a^3\equiv \mu$. Then $\delta h$
is computed in terms of $\delta a$ so that $\delta n=0$. Finally,
one must make use of the original radial equation (\ref{atwoa})
to determine $\delta a$.

We are especially interested in the case where the perturbing
potential $R$ is the sum of the quadrupole tide $R_2$,
Eq.~(\ref{2twentysixb}), and of a term with frequency $q=1$:
\begin{equation}
 R =R_2 + R_p = n'^2r^2\left[{1\over 4}+{3\over 4}\cos 2\left(v-v'
 \right)\right] + A r^p \cos\left(v-v'\right)\; .\label{anine}
\end{equation}
\narrowtext

In the case of the equivalence-principle-violating term
(\ref{2twentysixa}) the perturbing term in $\cos(v-v')$ has $p=1$,
while the octupolar tide (\ref{2twentysixc}), for which $p=3$,
contains a perturbing term in $\cos(v-v')$ (that we focus on) and
a term in $\cos 3(v-v')$.
At the fractional order ${\cal O}(m^2)$ beyond the first-order
solution (\ref{aeight}), the $\cos 3(v-v')$ term mixes with the
quadrupole tides $\propto \cos 2(v-v')$ to generate the frequency
$q=1$. As we work here only at the fractional order ${\cal O}(
 m)$ beyond the first-order solution, we do not need to study the
effect of the $\cos 3(v-v')$ term. The first-order solution
corresponding to Eq.~(\ref{anine}) reads
\begin{mathletters}
 \label{aten}
\begin{eqnarray}
 r &=& a + \delta^0 r + \delta^2 r + \delta^1 r \; , \label{atena}\\
 v &=& nt + \epsilon + \delta^2 v + \delta^1 v \; , \label{atenb}
\end{eqnarray}
\end{mathletters}
where the superscripts on $\delta$ indicate the value of the
frequency $q$. To sufficient accuracy for our purpose we have from
Eqs.~(\ref{aeight})
\begin{mathletters}
 \label{aeleven}
\begin{eqnarray}
  \delta^0 r+ \delta^2 r &=& -m^2a\left({1\over 6}+\cos 2\tau
 \right)+ {\cal O}(m^3) \; , \label{aelevena}\\
 \delta^2 v&=& {11\over 8} m^2 \sin 2\tau + {\cal O}(m^3)
 \; , \label{aelevenb}\\
 \delta^1 r &=& C^{\rm first-order}_r {A\over (n-n')^2} a^{p-1}
 \cos\tau \; , \label{aelevenc}\\
  \delta^1 v &=& C^{\rm first-order}_v {A\over (n-n')^2} a^{p-2}
 \sin\tau \; , \label{aelevend}
\end{eqnarray}
\end{mathletters}
with
\widetext
\begin{mathletters}
 \label{atwelve}
\begin{eqnarray}
 C^{\rm first-order}_r &=& {p+2+2m\over 2m + m^2}
 = {p+2\over 2m}\left[1+{2-p\over 2(2+p)}m+{\cal O}
 \left(m^2\right)\right] \; , \label{atwelvea} \\
  C^{\rm first-order}_v &=& -2 {(p+2+2m)(1+m)\over
 2m + m^2} +1
 = -2{p+2\over 2m}\left[1+{4+p\over 2(2+p)}m+{\cal O}
  \left(m^2\right)\right] \; , \label{atwelveb}
\end{eqnarray}
\end{mathletters}
\narrowtext
\noindent Note that the constant term $-m^2a/6$ in
Eq.~(\ref{aelevena})
depends on $\delta a=a_c-a$ and must be determined by having
recourse to Eq.~(\ref{atwoa}). Note also that the small denominator
present when $q=1$, $n^2-(n-n')^2 = (n-n')^2 [(1+m)^2-1]=
(n-n')^2 [2m + m^2]$, causes the synodic effects to be of
order ${\cal O}(m^{-1}A)$ instead of the usual order ${\cal O}(A)$
valid when $q\neq 1$. One of the effects of this small denominator
is to have a simple, approximate link between the radial and
longitudinal synodic oscillations: $ C^{\rm first-order}_v = -2
C^{\rm first-order}_r [1+{\cal O}(m)]$. Another effect is that
only the {\it leading} terms in Eqs.~(\ref{atwelve}) are correct. The
${\cal O}(m)$ fractional corrections are modified by higher
iterations as we are going to see.

When proceeding to the next iteration, several effects must be taken
into account. On the left-hand side of Eq.~(\ref{athreea}) one must
keep the terms of order $(\delta r)^2$, while on the right-hand
side one must include the change $\delta P$ of the source term
$P(r,v-v')$ induced by the first-order solution (\ref{aeleven}).
This leads to the following equation for $\delta r = r-a$
\begin{mathletters}
 \label{athirteen}
\begin{eqnarray}
 a\left({d^2\over dt^2}+n^2\right)\delta r &=& P(r_0,v_0-v_0')+
 \delta P_{\rm eff} \; ,\label{athirteena} \\
 \delta P_{\rm eff} &=& \delta P + \left(n^2-{1\over 2}{d^2\over
 dt^2}\right) (\delta r)^2\; . \label{athirteenb}
\end{eqnarray}
\end{mathletters}

When computing the synodic effects with fractional accuracy $1+{\cal
O}(m)$, the computation of $\delta P$ is simplified by several
circumstances. Because of the amplification ${\cal O}(m^{-1}A
)$ of the first-order synodic effects, one finds that it is enough
to compute in Eq.~(\ref{afive}) the change of the first
contribution, $(p+2)R_p$, for $p=2$ and under the synodic variations
$\delta^1 r$, $\delta^1 v$. This yields
\widetext
\begin{eqnarray}
 \left(\delta P\right)_{\rm synodic} &=& 4 \left(\delta R_2
 \right)_{\rm synodic}
 = \left\{\left(\delta^1 r{\partial\over \partial r}+\delta^1 v
 {\partial\over \partial v}\right)\left[n'^2r^2\left(1+3\cos2\left(v
 -v'\right)\right)\right] \right\}_{\rm synodic} \nonumber \\
 &=& 11\,n'^2 a^2 B \cos\tau \left[1+{\cal O}\left(m\right)
 \right]\; , \label{afourteen}
\end{eqnarray}
\narrowtext
\noindent
where $B\equiv (2m)^{-1}(p+2)(n-n')^{-2} Aa^{p-2}$ denotes the
leading value of the fractional synodic range oscillation ($\delta^1
r/a \simeq B\cos\tau$, $\delta^1 v\simeq -2B \sin\tau$). We need
also to extract the synodic piece of $(\delta r)^2 = (\delta^0 r
+ \delta^1 r + \delta^2 r)^2$ coming from the mixing between
$\delta^0 r + \delta^2 r \simeq -m^2a (\case1/6+\cos 2\tau)$ and
$\delta^1 r \simeq a B\cos \tau$: $(\delta r)_{\rm synodic}^2 \simeq
 -\case4/3 a^2 m^2 B \cos\tau
[1+{\cal O}(m)]$. Finally, the synodic piece of the
second-order effective source term for de Pont\'ecoulant's radial
equations (\ref{athirteen}) is obtained as
\begin{eqnarray}
 \delta^1 P_{\rm eff}&=& \left(\delta P\right)_{\rm synodic}+
 \left[n^2 + {1\over 2}\left(n-n'\right)^2\right] \left(\delta
 r^2\right)_{\rm synodic} \nonumber \\
 &=& 9 \,n'^2a^2 B\cos\tau \left[1+{\cal O}\left(m\right)
 \right]\; . \label{afifteen}
\end{eqnarray}

The corresponding solution reads
\begin{eqnarray}
 \left({\delta^1 r\over a}\right)^{\rm second-order} &\simeq&
 9{n'^2 \over n^2-(n-n')^2} B\cos\tau \nonumber\\
 &\simeq& {9\over 2} mB\cos\tau\; . \label{asixteen}
\end{eqnarray}

When turning to the longitude equation (\ref{athreeb}) one finds
also some simplifications: the change of the source term $\propto
\int dt\, \partial R/\partial v$ is of order $m^2 \times
m^{-1}A = {\cal O}(mA)$. The corresponding term in the
solution is not amplified by a small denominator and is therefore
negligible compared to the precision $m\times m^{-1}A
= {\cal O}(A)$ we are aiming for. It is therefore sufficient to
integrate the equation $d/dt\, (\delta^1 v)^{\rm second-order} \simeq
-2na^{-1}(\delta^1 r)^{\rm second-order}$. The final result can be
very simply expressed in saying that the second iteration leads
to multiplying the first-order synodic perturbations
(\ref{aelevenc}), (\ref{aelevend}) by the common factor $1+
\case9/2 m+  {\cal O}(m^2)$.

In conclusion, the mixing between the quadrupole tide $R_2$ and
some synodic-frequency perturbation potential (in which we factorize
an effective gravitational mass of the Sun, $Gm' \equiv n'^2 a'^3$)
\begin{equation}
 \left(R_p\right)_{\rm synodic} = \beta {Gm'\over a'} \left({r\over
 a'}\right)^p \cos\left(v-v'\right)\; , \label{aseventeen}
\end{equation}
leads, when neglecting non-linearities in the dimensionless
parameter $\beta$ (which should not be confused with its
post-Newtonian homonym), to the following synodic oscillations
\begin{mathletters}
 \label{aeighteen}
\begin{eqnarray}
 {\delta^1 r\over a} &=& C_r \beta m^2 \left({a\over a'}
 \right)^{p-2} \cos\tau\; , \label{aeighteena}\\
 \delta^1 v &=& C_v \beta m^2 \left({a\over a'}\right)^{p-2}
 \sin\tau \; , \label{aeighteenb}
\end{eqnarray}
\end{mathletters}
where
\begin{mathletters}
 \label{anineteen}
\begin{eqnarray}
 C_r &=& \left[1+{9\over 2} m + {\cal O}\left(m^2
 \right) \right] C_r^{\rm first-order} \nonumber \\
 &=& {2+p\over 2m}\left[1+{10+4p\over 2+p}m +
 {\cal O}\left(m^2\right) \right] \; , \label{anineteena} \\
 C_v &=& \left[1+{9\over 2} m + {\cal O}\left(m^2
 \right) \right] C_v^{\rm first-order} \nonumber \\
 &=& -2 {2+p\over 2m}\left[1+{11+5p\over 2+p}m +
 {\cal O}\left(m^2\right) \right] \; . \label{anineteenb}
\end{eqnarray}
\end{mathletters}

The two cases of interest are: (i) an hypothetical violation of the
equivalence principle in which (comparing Eq.~(\ref{aseventeen}) with
Eq.~(\ref{2twentysixa}))
\begin{equation}
 p=1\; ,\quad \beta_{\rm EP} ={\bar \delta}_{12}\; , \label{atwenty}
\end{equation}
and, (ii) the octupolar tide (``parallactic effects''),
Eq.~(\ref{2twentysixc}), with
\begin{eqnarray}
 R_3 &\simeq& {1\over 8}\left(X_2-X_1\right){Gm'\over a'}\left(
 {r\over a'}\right)^3 \nonumber \\
 & &\times\left[3\cos\left(v-v'\right)+5\cos3\left(v-v'\right)
 \right]\; ,
\end{eqnarray}
whose synodic piece has
\begin{equation}
 p=3\; ,\quad \beta_{\rm PAR} ={3\over 8}\left(X_2-X_1\right)\; .
 \label{atwentyone}
\end{equation}

The results (\ref{aeighteen})--(\ref{atwentyone}) agree with the
(much more accurate) Hill-Brown-type results given in the text and
in Appendix~B.

\section{ Hill's equations -- more on the iterative scheme of solution
 and numerical results}

In this Appendix, we explain in detail the iterative
scheme we employed for solving Hill's equations (\ref{2fourtyseven})
with the source terms (\ref{2fourtysix}), and also with the
parallactic perturbation (\ref{2fourtyeight}). We also give tables of
the obtained solution for several physically interesting quantities.
Obviously, one can envisage several iterative methods for solving
the considered equations. We do not claim that the scheme we adopted
is the optimal one, but we found it suitable from the point of
view of memory and computing time requests. Thanks to our use of the
dedicated algebraic manipulator MINIMS we could obtain the solution
to a very high order in the formally small parameters.
In what follows, we shall present the solution for the perturbation
of Hill's variational orbit related to the
equivalence principle violation. Exactly the same scheme applies in
the case of the parallactic perturbations.

Keeping the notation of Sec.~II.D, notably $L(w,{\bar w})$ for the
linear operator (\ref{2fourtysix}), we have to solve
\begin{eqnarray}
 L\left(w,{\bar w}\right) &=& -{\hat \lambda}\zeta^{-1} -{3\over 2}
 m^2 \zeta^{-2} \nonumber \\
 & & -{3\over 2}m^2 {\bar w}\, \zeta^{-2} + \kappa\left(
 m\right)Q\left(w,{\bar w}\right)\; , \label{bone}
\end{eqnarray}
where
\begin{equation}
 Q\left(w,{\bar w}\right)= \left(1+w\right)^{-1/2} \left(1+{\bar w}
 \right)^{-3/2} -1 +{1\over 2}w + {3\over 2}{\bar w}\; .
 \label{btwo}
\end{equation}
The nonlinear function $Q$ can be written as
\begin{eqnarray}
 Q\left(w,{\bar w}\right) & =& \sum_{k\geq 2} \chi_k\left(w^k +
  \left(2k+1\right){\bar w}^k\right) \nonumber \\
 & & +\sum_{j,k\geq 1} \left(2k+1\right) \chi_j\chi_k w^j {\bar w}^k
 \; . \label{bthree}
\end{eqnarray}
Here, $\chi_k \equiv \pmatrix{-1/2\cr k\cr}$ and $(2k+1)\chi_k =
\pmatrix{-3/2\cr k\cr}$ are binomial coefficients.

As in Sec.~II.D we look for a formal power series solution of
Eq.~(\ref{bone}),
\begin{equation}
 w = w^{(1)}+w^{(2)}+w^{(3)}+\cdots \; , \label{bfour}
\end{equation}
and similarly for the complex conjugate. The superscripts on the
consecutive terms in (\ref{bfour}) refer to corresponding orders
in the combined formal small parameter $({\hat \lambda} + m^2)$.
Keeping track of the orders in this formal small parameter, we
decompose the nonlinear source $Q$ as
\begin{equation}
 Q\left(w,{\bar w}\right) = Q^{(2)} + Q^{(3)} + \cdots \; ,
 \label{bfive}
\end{equation}
where the individual terms include symbolically
\begin{equation}
 Q^{(i)} = \sum_{ja+kb = i} ({\rm coefficient})
 \,\left(w^{(a)}\right)^j \left({\bar w}^{(b)}\right)^k\; .
 \label{bsix}
\end{equation}
For any particular value of $i$ in (\ref{bsix}), $Q^{(i)}$ is given
by a finite number of terms which depend only on the knowledge of
$w^{(a)}$ for $a<i$. Although the procedure of
breaking $Q(w,{\bar w})$ into a sum of equal-order terms $Q^{(i)}$
might seem laborious, it is relatively easy to be
programmed using a well suited algebraic manipulator such as MINIMS.
One can introduce a formal index which conserves the order of a
particular term and manipulate it like any other variable.

The heart of our iteration scheme consists of the following infinite
system of differential equations
\widetext
\begin{eqnarray}
 L\left(w^{(1)},{\bar w}^{(1)}\right) &=& -{\hat \lambda}\zeta^{-1}
 -{3\over 2} m^2 \zeta^{-2} \; ,  \nonumber \\
 L\left(w^{(2)},{\bar w}^{(2)}\right) &=& -{3\over 2}m^2
 {\bar w}^{(1)}\zeta^{-2} + \kappa
 Q^{(2)}\left(w^{(1)}\right) \; , \nonumber \\
 L\left(w^{(3)},{\bar w}^{(3)}\right) &=& -{3\over 2}m^2
 {\bar w}^{(2)}\zeta^{-2} + \kappa
 Q^{(3)}\left(w^{(1)},w^{(2)}\right) \; , \nonumber \\
  \ldots & = & \ldots \; , \nonumber \\
 L\left(w^{(k)},{\bar w}^{(k)}\right) &=& -{3\over 2}m^2
 {\bar w}^{(k-1)}\zeta^{-2} +  \kappa
 Q^{(k)}\left(w^{(1)},w^{(2)},\ldots,w^{(k-1)}\right) \; ,  \nonumber \\
  \ldots & = & \ldots \; . \label{bseven}
\end{eqnarray}
\narrowtext

It is easy to verify, that the generic form of the terms in the
right-hand sides of Eqs.\ (\ref{bseven}) reads $W_k\zeta^k+W_{-k}
\zeta^{-k}$ as presented in (\ref{2fifty}). The unique inversion
of the linear operator $L$ on the left-hand sides of Eqs.\
(\ref{bseven}) is given by formulas
(\ref{2fiftyone})--(\ref{2fiftytwo}). Notice also that suppressing
the equivalence-principle-violation term (${\hat \lambda}=0$ in
(\ref{bseven})), one recovers a system of equations for constructing
the usual variational periodic orbit.

In the preceding scheme, we consider ${\hat \lambda}$ and $m^2$ as
two comparable ``small'' parameters. However, in practice,
the order of the ${\hat \lambda}$ parameter associated with the
studied violation of the equivalence principle is numerically
much smaller than $m^2$ (which can be as large as $1$).
As a result, we restrict the generality of our solution by keeping
only {\it the first order} in the parameter ${\hat \lambda}$.
This truncation allows a clear separation
in the interpretation of the odd- and even-power terms in the
$\zeta$ variable of the final solution for $w$ (and ${\bar w}$):
(i) the even-power terms ($\propto \zeta^{2i}$) never contain the
perturbation parameter ${\hat \lambda}$ and fully reconstruct
Hill's variational solution, (ii) the odd-power terms ($\propto
\zeta^{2i+1}$) are all of the first-order in
${\hat \lambda}$ (but they are coupled to the
``background'' variational terms through an infinite series of powers
of $m^2$). We thus {\it simultaneously} obtain Hill's variational
solution and its ${\hat \lambda}$-perturbation by filtering the
various powers of $\zeta$. This is a
particularly important circumstance, because
the series giving the variational solution enters the definition of
several studied quantities such as the radial or longitudinal
perturbations of the lunar orbit
by the equivalence-principle-violation terms (see Eqs.\
(\ref{2fiftyeight})).

Once the iterative scheme is set up and the numerical program
debugged, we can obtain the solution of our problem up to an
arbitrary order. The limits of the solution are then
given mainly by the computer power. A minor limit comes from
the fact that the MINIMS algebraic manipulator
works with double precision real coefficients ($16$ digits)
\cite{Moons}. During the manipulation of the series, one
thereby accumulates round-off errors.
However, we have checked that this restriction is
not significant for our work\footnote{ We have performed
a lower order solution in using the modified version of the
distributed MINIMS manipulator which accepts the quadruple
precision ($32$ digits) for the coefficients of the series
and compared it with the double precision one.}.

In the rest of this Appendix we shall present tables of the
numerical coefficients achieved by the previous algorithm for
different series introduced in the main text of the paper and
related to physical quantities.

Let us start with our solution for the variational curve
$({\hat \lambda} =0)$. Tables I and II give the coefficients
$w_{ij}$ of the double series expansion of $w$:$w=\sum w_{jk}
\zeta^j m^k$ where $j=0,\pm 2,\pm 4,\ldots$; $k=2,3,4,\ldots$
and $\vert j \vert \leq k$ ($j$ labels the rows and $k$ the
columns). Contrary to the method of Ref. \cite{BC}
the $\zeta$-independent
term is not fixed to unity. However, due to the choice of the fiducial
semimajor axis ${\tilde a}$, defined in Eq.~(\ref{2thirtysix}), it
starts only at the power $m^4$.

\widetext
\begin{table}
\squeezetable
\caption{ Coefficients $w_{jk}$ of the double series giving Hill's
variational curve: $w=\sum w_{jk} \zeta^j m^k$; $j$ labels the rows
and $k$ the columns.}
\begin{tabular}{rccccc}
 & $2$ & $3$ & $4$ & $5$ & $6$ \\
\tableline
-6 & -- & -- & -- & -- & 0.005208333333333333 \\
-4 & -- & -- & 0.0000000000000000 & 0.03593750000000000 &
     0.1245833333333333 \\
-2 & -1.187500000000000 & -1.666666666666667 & -1.194444444444444 &
     -0.5185185185185185 & -0.8265365788966050 \\
0  & -- & -- & 0.6210937500000000 & -0.6770833333333333 &
     -0.6892361111111111 \\
 2 & 0.1875000000000000 & 0.5000000000000000 & 0.5833333333333333 &
     0.3055555555555556 & -0.1615849247685185 \\
 4 & -- & -- & 0.0976562500000000 & 0.4182291666666667 &
     0.8484722222222222 \\
 6 & -- & -- & -- & -- & 0.06778971354166667  \\
\end{tabular}
 \label{tabt}
\end{table}

\begin{table}
\squeezetable
\caption{ Continuation of Table~I.}
\begin{tabular}{rccccc}
 & $7$ & $8$ & $9$ & $10$ & $11$ \\
\tableline
-10 & -- & -- & -- & 0.002728271484375000 & 0.02314830588154192 \\
-8  & -- & 0.003743489583333333 & 0.02753602458235899 &
      0.09640929637552359 & 0.2138120808825399 \\
-6  & 0.03477027529761905 & 0.1040162627551020  & 0.1880280591858106 &
      0.2403927412241086 & 0.2416131234195372 \\
-4  & 0.1956215277777778 & 0.2153863326461227  & 0.2955465183221726 &
      0.3769043026659738 & -0.1199168976723571 \\
-2  &  1.205252137988683 &  7.122740269204389  & 15.70605739258259  &
      19.78110897697562  & 14.66018287215882 \\
0   & -1.605902777777778 & -0.5700574333285108 & 0.3975170245386445 &
      1.670582431118184  & 7.457045029767965 \\
 2  & -1.034723427854938 & -3.070661490483539  & -6.516200630679869 &
      -9.530344734021654 & -7.910677339637028 \\
 4  &  1.038887731481481 & 0.7041097601996523  & -0.5710030937052194 &
      -4.063905605879704 & -11.57027457856626 \\
 6  & 0.3898297991071429 &  1.087330552012472  & 1.926893073875549  &
       2.301395728392179 &  1.149297956477343 \\
 8  & -- & 0.05397033691406250 & 0.3869401996638499 & 1.360005033209030 &
       3.095895579274965 \\
 10 & -- & -- & -- & 0.04656556447347005 & 0.3988932608444074 \\
\end{tabular}
 \label{tabt1}
\end{table}
\narrowtext

We then give the lunar orbit
perturbations due to a hypothetical equivalence principle violation
(terms linear in $\hat \lambda$).

Tab.~\ref{tab1} gives the coefficients $c_k$ of the series in powers
of $m$ giving $w_{-1}$, i.e. the coefficient of $\zeta^{-1}$ in
the Laurent expansion of $\delta_{\lambda} w(\zeta)$, after factorization
of the leading term $\case9/4 ({\hat \lambda}/m)$ (see
Eq.~(\ref{2fiftysixb})). The second column --
$p_k$ -- gives the numerical value of $c_k m^k$ (in \%) for the lunar
orbit ($m=m_{\rm Moon} =0.0808489375\ldots$). The last column --
$r_k$ -- gives the ratio $c_{k-1}/c_k$ of the successive coefficients
of the series (the same structure is conserved also for Tables
\ref{tab2}--\ref{tab3} and \ref{tab5}--\ref{tab6}).
\begin{table}
\caption{Coefficients of $m^k$ in the lunar orbit perturbation
$w_{-1}$. The mark $<$ signifies that the value is smaller than
$0.001$ \%.}
\begin{tabular}{rccc}
 $k$ & $c_k$ & $p_k$ & $r_k$ \\
\tableline
0  & 1.000000000000000 & --     & --           \\
1  & 5.111111111111111 & 41.323 & 0.1956521739 \\
2  & 25.52777777777778 & 16.686 & 0.2002176279 \\
3  & 129.5277777777778 &  6.845 & 0.1970834227 \\
4  & 663.1076388888889 &  2.833 & 0.1953344679 \\
5  & 3400.509837962963 &  1.175 & 0.1950024174 \\
6  & 17434.56978202160 &  0.487 & 0.1950440923 \\
7  & 89366.97811374742 &  0.202 & 0.1950896198 \\
8  & 458049.7173103071 &  0.084 & 0.1951032273 \\
9  & 2347711.432406117 &  0.035 & 0.1951047778 \\
10 & 12033105.92287766 &  0.014 & 0.1951043602 \\
11 & 61675313.45886662 &  0.006 & 0.1951040902 \\
12 & 316115045.4115870 &  0.002 & 0.1951040115 \\
13 & 1620238694.871019 &  0.001 & 0.1951039970 \\
14 & 8304487501.579596 &  $<$   & 0.1951039958 \\
15 & 42564415192.16274 &  $<$   & 0.1951039962 \\
16 & 218162702733.8958 &  $<$   & 0.1951039965 \\
17 & 1118186744062.939 &  $<$   & 0.1951039966 \\
\end{tabular}
 \label{tab1}
\end{table}

Tab.~\ref{tab1b} gives the coefficients $c_k$ of the series in powers of
$m$ giving $w_1$, i.e. the coefficient of $\zeta$ in the
Laurent expansion of $\delta_{\lambda} w$, after factorization
of the leading term $-\case3/4
({\hat \lambda/m})$ (see Eq.~(\ref{2fiftysixa})).
\begin{table}
\caption{ Coefficients of $m^k$ in the lunar orbit perturbation $w_{1}$.
Notation as in Table~III.} \begin{tabular}{rccc}
 $k$ & $c_k$ & $p_k$ & $r_k$ \\
\tableline
0  & 1.000000000000000 & --     & --           \\
1  & 6.000000000000000 & 48.509 & 0.1666666667 \\
2  & 29.62500000000000 & 19.365 & 0.2025316456 \\
3  & 147.5000000000000 &  7.795 & 0.2008474576 \\
4  & 749.9427083333333 &  3.204 & 0.1966816910 \\
5  & 3843.245659722222 &  1.328 & 0.1951326495 \\
6  & 19711.47459129051 &  0.551 & 0.1949750457 \\
7  & 101057.1821729118 &  0.228 & 0.1950526837 \\
8  & 517989.9317152261 &  0.095 & 0.1950948773 \\
9  & 2654938.425605692 &  0.039 & 0.1951043108 \\
10 & 13607759.81112155 &  0.016 & 0.1951047389 \\
11 & 69746091.35545381 &  0.007 & 0.1951042639 \\
12 & 357481499.9590677 &  0.003 & 0.1951040581 \\
13 & 1832261194.517921 &  0.001 & 0.1951040065 \\
14 & 9391202741.857013 &  $<$   & 0.1951039973 \\
15 & 48134343317.40093 &  $<$   & 0.1951039963 \\
16 & 246711211454.4727 &  $<$   & 0.1951039964 \\
17 & 1264511316090.325 &  $<$   & 0.1951039966 \\
\end{tabular}
 \label{tab1b}
\end{table}

Some general properties of the presented results, already commented upon
in the main text of the paper and in Appendix~C, are: (i) a
fast (``geometric-like'') growth of
the coefficients $c_k$ of the series with a surprisingly rapid
approach of the $r_k(=c_{k-1}/c_k)$ ratio to the $m_{cr}$ value
of about $0.1951039966\ldots$, (ii) a very substantial contribution
of the non-linear terms ($k\geq 1$) in the series for the lunar value
of Hill's parameter $m_{\rm Moon}$.
\begin{table}
\caption{ Coefficients of the $S(m)$ series yielding the
 radial perturbation, with synodic (``$\tau$'') period, of
 the variational curve due to an equivalence principle violation.}
\begin{tabular}{rccc}
 $k$ & $c_k$ & $p_k$ & $r_k$ \\
\tableline
0  &   1.000000000000000 &  --     & --            \\
1  &   4.666666666666667 &  37.730 & 0.21428571429 \\
2  &   22.10416666666667 &  14.449 & 0.21112158341 \\
3  &   111.0416666666667 &   5.868 & 0.19906191370 \\
4  &   569.0742187500000 &   2.431 & 0.19512686221 \\
5  &   2920.624565972222 &   1.001 & 0.19484675483 \\
6  &   14976.54921694155 &   0.418 & 0.19501318519 \\
7  &   76767.66017493731 &   0.173 & 0.19508930170 \\
8  &   393469.7706768071 &   0.072 & 0.19510434065 \\
9  &   2016707.919972300 &   0.030 & 0.19510498609 \\
10 &   10336561.13767503 &   0.012 & 0.19510433819 \\
11 &   52979731.33709500 &   0.005 & 0.19510406861 \\
12 &   271546096.6185635 &   0.002 & 0.19510400627 \\
13 &   1391801807.779853 &   0.001 & 0.19510399764 \\
14 &   7133640668.991231 &   $<$   & 0.19510399701 \\
15 &   36563272849.28028 &   $<$   & 0.19510399680 \\
16 &   187404017729.6833 &   $<$   & 0.19510399666 \\
17 &   960533976540.2227 &   $<$   & 0.19510399664 \\
\end{tabular}
 \label{tab2}
\end{table}
\begin{table}
\caption{ Coefficients of the $S'(m)$ series yielding the
 longitudinal perturbation, with synodic (``$\tau$'') period,
 of the variational curve due to an equivalence principle violation.}
\begin{tabular}{rccc}
 $k$ & $c_k$ & $p_k$ & $r_k$ \\
\tableline
0  &  1.000000000000000 & --     & --            \\
1  &  5.333333333333333 & 43.119 & 0.18750000000 \\
2  &  26.39583333333333 & 17.254 & 0.20205209155 \\
3  &  132.9166666666667 &  7.024 & 0.19858934169 \\
4  &  678.3172743055553 &  2.898 & 0.19595058493 \\
5  &  3478.588686342593 &  1.202 & 0.19499783834 \\
6  &  17838.85192117573 &  0.498 & 0.19500070418 \\
7  &  91447.83453394253 &  0.206 & 0.19507134326 \\
8  &  468721.9555645895 &  0.086 & 0.19510038616 \\
9  &  2402405.984973695 &  0.035 & 0.19510522305 \\
10 &  12313419.50072995 &  0.015 & 0.19510469735 \\
11 &  63112018.54675127 &  0.006 & 0.19510419385 \\
12 &  323478806.4318246 &  0.003 & 0.19510402936 \\
13 &  1657981435.159388 &  0.001 & 0.19510399789 \\
14 &  8497936880.202027 &  $<$   & 0.19510399507 \\
15 &  43555934597.05303 &  $<$   & 0.19510399579 \\
16 &  223244707500.3778 &  $<$   & 0.19510399635 \\
17 &  1144234415604.679 &  $<$   & 0.19510399657 \\
\end{tabular}
 \label{tab3}
\end{table}

Tables~\ref{tab2}--\ref{tab3} give the coefficients of the
power series $S(m)$ and $S'(m)$ related, respectively,
to the radial and longitudinal perturbations (with synodic
periodicity) of the lunar orbit due to a hypothetical violation of
the equivalence principle. The coefficients of the Pad\'e
approximants of those series are given in Tab.~\ref{tab4}. More
precisely, we denote $N_8(m) \equiv \sum_0^8 a_k m^k$
and $D_8(m) \equiv \sum_0^8 b_k m^k$ the Pad\'e approximants
for the $S(m)$ series (see Eq.\ (\ref{3two})),
and $N'_8(m) \equiv \sum_0^8 a'_k m^k$ and $D'_8(
m) \equiv \sum_0^8 b'_k m^k$ the Pad\'e approximants
for the $S'(m)$ series. The denominator polynomials $D_8(
m)$ and $D'_8(m)$ have $0.1951039967\dots$ and $0.1951039966\ldots$,
respectively, as real roots.
\widetext
\begin{table}
\caption{ Coefficients of the Pad\'e approximants of order $8$ of the
 radial and longitudinal perturbation series, with synodic (``$\tau$'')
 period, of the variational curve due to an
 equivalence principle violation.}
\begin{tabular}{rcccc}
 $k$ & $a_k$ & $b_k$ & $a'_k$ & $b'_k$ \\
\tableline
0 &   1.000000000000000    &   1.000000000000000 &
      1.000000000000000    &   1.000000000000000   \\
1 &  -4.679659654425508    &  -9.346326321092175 &
     -2.948114664716526    &  -8.281447998049859   \\
2 &   9.706375765630793    &  31.21839859739427  &
      2.472472675594219    &  20.24436199852680    \\
3 & -12.09649460259615     & -62.23126666796116  &
     -0.035120215085467699 & -22.32599642537073    \\
4 &   9.264485113438459    &  78.37781038863183  &
     -0.8074761600241968   &   6.322888291078106   \\
5 &  -0.8206525542778040   & -59.42770492795888  &
      1.129701196880901    &   4.768342316076093   \\
6 &  -4.562346877625465    &   5.529574170471975 &
      1.916287015273362    &  13.88318783117331    \\
7 &   0.6450458113816286   &  30.28114298289518  &
     -3.678767273115525    & -38.01361281338056    \\
8 &   0.9269455883562498   & -30.33613801897244  &
     -3.906968891383218    &  34.97077854522613    \\
\end{tabular}
 \label{tab4}
\end{table}
\narrowtext

Tab.~\ref{tab5} gives the coefficients of the $S^{\prime\prime}(
m)$ series yielding the equivalence-principle-violation
perturbation of the lunar orbit with one-third of the synodic
period (``$3\tau$''). This series is defined by
\begin{equation}
 \left(\delta_\lambda r\right)_{\rm third-synodic} = C^{\prime
 \prime}\left(m\right) {\bar \delta}_{12} a' \cos 3\tau\; ,
 \label{beight}
\end{equation}
with
\begin{equation}
 C^{\prime\prime}\left(m\right)= {51\over 32}m^3 \left(1
 +\sum_{k\geq 1}c^{\prime\prime}_k m^k\right)\equiv {51\over 32}
 m^3\, S^{\prime\prime}\left(m\right)\; . \label{bnine}
\end{equation}
Note that the third-synodic effect (\ref{beight}) is ${\cal O}(
m^2)$ smaller than the synodic effect (\ref{2sixty}). For instance,
if we assume ${\hat \delta}_{12}=0$ the result (\ref{bnine})
gives numerically $(\delta_\lambda r)_{\rm
third-synodic} \simeq 6.62 \eta \cos 3\tau$ centimeters for the
lunar orbit. Although there is probably no practical use of this
higher frequency excitation (and of the others with frequencies
$5\tau$, $7\tau$, $\ldots$), two points are to be mentioned:
(i) a significantly smaller amplitude of the effect,
(ii) the persistence of the pole singularity at $m=m_{cr}$
for these odd-multiples of the basic synodic frequency as discussed
in Appendix~C.
\begin{table}
\caption{ Coefficients of the $S^{\prime\prime}(m)$ series
 yielding the radial perturbation, with third-synodic (``$3\tau$'')
 period, of the variational curve due to an
 equivalence principle violation.}
\begin{tabular}{rccc}
 $k$ & $c^{\prime\prime}_k$ & $p_k$ & $r_k$ \\
\tableline
1  &  1.000000000000000 & --     & --            \\
2  &  6.450980392156863 & 52.155 & 0.15501519757 \\
3  &  32.60457516339869 & 21.312 & 0.19785506665 \\
4  &  163.0646514161220 &  8.618 & 0.19994876192 \\
5  &  828.6388478122731 &  3.540 & 0.19678615340 \\
6  &  4240.758469846889 &  1.465 & 0.19539873674 \\
7  &  21736.94075865620 &  0.607 & 0.19509454053 \\
8  &  111428.4766430605 &  0.252 & 0.19507527531 \\
9  &  571159.5105168029 &  0.104 & 0.19509169434 \\
10 &  2927510.871437882 &  0.043 & 0.19510073083 \\
11 &  15004916.03283528 &  0.018 & 0.19510344910 \\
12 &  76907280.99332377 &  0.007 & 0.19510397246 \\
13 &  394186033.1895487 &  0.003 & 0.19510402327 \\
14 &  2020389145.708269 &  0.001 & 0.19510401450 \\
15 &  10355446785.91400 &  $<$   & 0.19510400541 \\
16 &  53076547791.36041 &  $<$   & 0.19510400011 \\
17 &  272042338525.2278 &  $<$   & 0.19510399771 \\
\end{tabular}
 \label{tab5}
\end{table}

Tab.~\ref{tab6} lists the coefficients of the $Q(m)$ series
giving the radial parallactic inequality of the lunar motion ($Q(
m)\equiv 1+ \sum_{k\geq 1} q_k m^k$) as defined in Eqs.~(\ref{3four})
of the text. Similarly to the
treatment of the equivalence-principle-violation lunar perturbation
we improved on our solution by using Pad\'e approximants.
Tab.~\ref{tab7} yields the coefficients of the corresponding
polynomials.
We also computed the corresponding lunar parallactic inequality in
longitude. As a partial check on our results we have compared the
latter with the result by Deprit, Henrard and Rom
\cite{DHR}. When substituting the current recommended values of the
mass constants and the
Earth semi-major axis we obtain $125."438$ for the amplitude
of the parallactic inequality in longitude when truncating our
series to the power $m^7$. This value is to be compared with
$125."4201$ reported in Ref.~\cite{DHR}. We believe that the
origin of the minor discrepancy between those results lays in
the slightly different values of the astronomical constants employed
by Deprit, Henrard and Rom at the end of the sixties.
\begin{table}
\caption{ Radial perturbation of the variational curve due to
 parallactic terms with synodic (``$\tau$'') period.}
\begin{tabular}{rccc}
 $k$ & $q_k$ & $p_k$ & $r_k$ \\
\tableline
0  & 1.000000000000000 & --     & --           \\
1  & 4.400000000000000 & 35.574 & 0.2272727273 \\
2  & 13.43750000000000 &  8.783 & 0.3274418605 \\
3  & 59.99895833333333 &  3.171 & 0.2239622216 \\
4  & 318.5420138888889 &  1.361 & 0.1883549288 \\
5  & 1665.565227141204 &  0.575 & 0.1912515996 \\
6  & 8563.762388478974 &  0.239 & 0.1944898926 \\
7  & 43904.44785527987 &  0.099 & 0.1950545516 \\
8  & 225048.4315368017 &  0.041 & 0.1950888862 \\
9  & 1153503.664577334 &  0.017 & 0.1950998843 \\
10 & 5912167.013677382 &  0.007 & 0.1951067454 \\
11 & 30302311.46598434 &  0.003 & 0.1951061397 \\
12 & 155313062.3650712 &  0.001 & 0.1951047195 \\
\end{tabular}
 \label{tab6}
\end{table}

\widetext
\begin{table}
\caption{ Coefficients of the Pad\'e approximants of order $6$ of the
 radial and longitudinal perturbations of the variational curve due to
 the parallactic terms with synodic (``$\tau$'') period.}
\begin{tabular}{rcccc}
 $k$ & $a^{\prime\prime}_k$ & $b^{\prime\prime}_k$ & $a^{\prime\prime
 \prime}_k$ & $b^{\prime\prime\prime}_k$ \\
\tableline
 0 &  1.000000000000000 &  1.000000000000000 &
      1.000000000000000 &  1.000000000000000 \\
 1 & -2.698270836202914 & -7.098270836202914 &
     -1.839002476253125 & -7.039002476253125 \\
 2 & -6.296177502041919 &  11.49871417725090 &
     -7.776514253041273 &  11.36379862347498 \\
 3 &  9.635571285915953 & -5.574715065844693 &
      4.712486681328506 & -6.420060419171178 \\
 4 &  11.54422294537151 & -11.09416027055608 &
      8.382766821277098 & -10.90642205141127 \\
 5 & -7.427911939834412 &  21.91801429175468 &
     -4.643821407637831 &  18.97128636175652 \\
 6 & -10.77769175983123 & -27.61496107523247 &
     -3.937875173153431 & -20.15607339834716 \\
\end{tabular}
 \label{tab7}
\end{table}
\narrowtext\noindent

\section{ Characteristic multipliers, commensurabilities and
 instability}

Let us first recall the basic concept of characteristic multipliers.
The small perturbations around a periodic orbit in the restricted
three-body problem can be described in terms of a two-dimensional
Poincar\'e map: this is the application connecting two successive
intersections of the trajectory in phase space with a two-plane
transversal to the orbit. [One works, say, in the three-dimensional
reduced phase space corresponding to a fixed value of the Jacobi
integral; see \cite{CM91} for a catalog of such Poincar\'e maps in
the case of the Hill problem.] For infinitesimal perturbations, the
Poincar\'e map reduces to a linear transformation of the plane,
leaving fixed the origin which corresponds to the reference
periodic orbit. The two eigenvalues $(\lambda_1,\lambda_2)$ of the
infinitesimal Poincar\'e map (a $2\times 2$ matrix) are the
characteristic multipliers. From the Hamiltonian nature of the
dynamics it follows that these multipliers are either of the form
$(e^{i\alpha},e^{-i\alpha})$ or $(\varepsilon e^\beta,
\varepsilon e^{-\beta})$ with $\varepsilon= \pm 1$ \cite{Arnold}. The
first case
means generally (apart from the exceptional cases where $\alpha =
2\pi/3$, or where $\alpha=2\pi/4$ and some inequality is not
satisfied) that the periodic orbit is (quasi-)stable. The second
case means that the periodic orbit is unstable. A useful
quantity for studying the possible loss of stability is half the sum
of the multipliers: $a\equiv \case1/2 (\lambda_1+\lambda_2)$, which
is either $\cos\alpha$ (in which case $|a| \leq 1$) or $\varepsilon
\cosh\beta$ (in which case $|a| \geq 1$). The loss of stability can
occur (apart from the above mentioned exceptional cases) only when
$a$ crosses the values $\pm 1$. The stability of all the periodic
orbits in Hill's problem has been studied by H\'enon \cite{Henon}.
We are interested here in the families ``{\it g}'' and ``{\it f}'' of
periodic orbits which correspond, respectively, to prograde and
retrograde lunar-type orbits. H\'enon found that retrograde orbits
are stable for all values of the Jacobi integral (i.e. $-1<m
<0$), while prograde orbits are stable only for close enough orbits,
$0<m<m_{cr}$ with\footnote{ In H\'enon's fourth paper \cite{Henon},
he gives the value $\pi m_{cr} = T/2 = 0.61294$. The more precise
value (\ref{cone}) was privately communicated to us by H\'enon, and
follows also independently from our results in Appendix B
(study of the geometric-like
growth of various $m$-series and of the zeroes of Pad\'e
denominators).}
\begin{equation}
 m_{cr} = 0.1951039966\ldots \; . \label{cone}
\end{equation}
For this value $a$ crosses the value $+1$.

Let us translate this result in terms of the perigee precession of
a perturbed Hill orbit. Perturbations of Hill's orbit can be
described in terms of the (isoenergetic) {\it normal} displacement
$q = ({\dot X}\delta Y-{\dot Y}\delta X)/({\dot X}^2+
{\dot Y}^2)^{1/2}$. This variable satisfies ``Hill's equation''
\begin{equation}
 {d^2q(\tau) \over d\tau^2}+ \Theta(\tau) q(\tau) = \sigma(\tau)
 \; , \label{ctwo}
\end{equation}
where $\Theta(\tau)=\theta_0 +2\sum_1^\infty \theta_j\cos 2j\tau$ is
periodic with period $\pi$. The source term $\sigma(\tau)$ on the
right-hand side of Eq.\ (\ref{ctwo}) is zero for free perturbations
(i.e. corresponding to adding some ``eccentricity'' to Hill's
``circular-like'' variational orbit), and non-zero when one perturbs
Hill's Hamiltonian $H_{\rm Hill}=({\rm kinetic\; terms})+F_0+F_2$
(e.g. by adding the $F_1$ perturbation we are mainly concerned with,
or the parallactic terms $F_3 + \cdots$). Perigee precession is
described by the general homogeneous solution of Eq.\ (\ref{ctwo})
($\sigma=0$). The latter general homogeneous solution can be
written as a linear combination of complex solutions of the form
$q(\tau) = \zeta^{\rm c} \sum_j b_j\zeta^{2j}$ and of their complex
conjugates (we recall that $\zeta=e^{i\tau}$). On the one hand,
the quantity ${\rm c}$ is linked to the usual rate of perigee advance
$d\varpi/dt$ (in the non-rotating frame) by \cite{BC}
\begin{equation}
 {\rm c} = {dl \over d\tau} = \left(1+m\right)\left(1-{1\over n}
 {d\varpi \over dt}\right) \label{cthree}
\end{equation}
(where $l=nt+\epsilon-\varpi$ is the ``mean anomaly''). On the other
hand, the quantity
${\rm c}$ is directly linked to the characteristic multipliers.
Indeed, when $\tau \rightarrow \tau+2\pi$, $q(\tau)$ gets multiplied
by $e^{2\pi i {\rm c}}$, while ${\bar q}(\tau)$ gets multiplied
by $e^{-2\pi i {\rm c}}$. Therefore the half-sum of the
multipliers is simply
\begin{equation}
 a=\cos(2\pi{\rm c})\; . \label{cfour}
\end{equation}

By using the perturbative series giving the perigee precession,
$n^{-1}(d\varpi/dt) = \case3/4 m^2+{177\over 32}m^3 +
\cdots$ (which has been computed to high accuracy in Ref.~\cite{DHR};
see also \cite{BI85}), one can check that the crossing of $a=+1$
found by H\'enon corresponds, when $m$ increases from $0$ to
$m_{cr}$, to ${\rm c}$ increasing from $1$ to a slightly
higher value ($\simeq 1.1$) and then decreasing to reach the value
$1$ at $m=m_{cr}$. From the smoothness of the variation
of the characteristic multipliers, and therefore of
$a$, with $m$ we deduce that, beyond $m=m_{cr}$, ${\rm
c}(m)$ goes through a quadratic branch point ${\rm c}(
m) \simeq (1-m/m_{cr})^{1/2}$ and becomes
complex\footnote{The combined facts that $m_{cr}$ is rather
small and that ${\rm c}(m)$ has a quadratic branch point at
$m=m_{cr}$ ``explains'' the notoriously bad convergence
of the perturbation series giving $d\varpi/dt$. Rewriting this
series in terms of better behaved quantities, such as $\cos(2\pi
{\rm c})$ or $\cos(\pi{\rm c})$, improves more its convergence
than the ``Euler transformations'', $m \rightarrow m/
(1+\alpha m)$, which have been traditionally used
\cite{Henrard}.}.

Finally, the important information for our purpose is that when
$m$ increases up to $m_{cr}$, the quantity ${\rm c}(
m) \simeq (1-m/m_{cr})^{1/2}$ is such that both
functions $\cos2\pi{\rm c}(m)\, (=a)$ and $\cos\pi{\rm c}({\rm
m})$ cross smoothly (without branch points or discontinuities
of derivatives) their corresponding limiting values $\cos2\pi=+1$
and $\cos\pi=-1$.

Let us now consider Hill's way of solving equation (\ref{ctwo}).
When inserting $q(\tau)=\zeta^{\rm c}\sum_j b_j\zeta^{2j}$ into
(\ref{ctwo}) one gets an infinite system of linear equations for
the coefficients $b_j$. When written in a suitably normalized way,
the determinant of this infinite system (which depends on ${\rm c}$),
say $\Delta({\rm c})$, is a well-defined quantity (Hill's
determinant). An homogeneous solution ($\sigma=0$) exists only for
the values of ${\rm c}$ for which $\Delta({\rm c})=0$. On the other
hand, if we consider the case where there is a source term on the
right-hand side of Eq.\ (\ref{ctwo}) of the form $\sigma(\tau)=
\zeta^{{\rm c}_\sigma}\sum_j \sigma_j\zeta^{2j}$, the corresponding
inhomogeneous solution $q_{\rm inhom}(\tau)$ will have the same
form as $\sigma(\tau)$, but will contain, as for usual finite
determinants, a factor $1/\Delta({\rm c}_\sigma)$. This factor will
become infinite when ${\rm c}_\sigma$ tends to one of the free
perigee precession values for which $\Delta({\rm c})=0$. The
analysis of the determinant $\Delta({\rm c})=0$ \cite{Brown},
\cite{BC} shows that, as a function of ${\rm c}$, it is a linear
function of $\cos\pi{\rm c}$. We conclude that the forced
perturbation $q_{\rm inhom}(\tau)$ will contain the factor $(\cos
\pi{\rm c}_\sigma - \cos\pi{\rm c}(m))^{-1}$, where ${\rm c}
(m)$ denotes the free perigee precession value. From our
analysis above, we know that when $m$ crosses $m_{cr}$,
$\cos\pi{\rm c}$ crosses smoothly $-1$. The final conclusion is that
the source terms for which $\cos\pi{\rm c}_\sigma=-1$, i.e. ${\rm
c}_\sigma =\pm 1,\pm 3,\pm 5, \ldots$, generates normal displacements
of Hill's orbit which have pole singularities $\propto (m-
m_{cr})^{-1}$ as $m$ crosses the value (\ref{cone}) found in
linear stability analyses. Moreover, it is easily checked that the
addition of a perturbing potential, say $F_p$, to Hill's potential
$F_0+F_2$, generates a source term $\sigma(\tau)$ in Hill's
variational equation (\ref{ctwo}) which is a linear combination of
$F_p$ and $\Re e\,[Du(\partial F_p/\partial u)]$ with real
coefficients of the form $k_0 + 2\sum_j k_j \cos 2j\tau$. More
precisely
\begin{equation}
 \sigma(\tau) = -2\varphi \left\{ \varphi^2\Phi F_p +\Re e\,\left[
 Du\left(\partial F_p/\partial u\right)\right]\right\} \; ,
 \label{cfive}
\end{equation}
where
\begin{mathletters}
 \label{csix}
\begin{eqnarray}
 \varphi^{-2}(u,{\bar u}) &=& - Du D{\bar u} \; , \label{csixa}\\
 \Phi(u,{\bar u}) &=& mDu D{\bar u}-2\Re e\,\left[Du\left(
 \partial F_{\rm Hill}/\partial u\right)\right]\; , \label{csixb}
\end{eqnarray}
\end{mathletters}
with $F_{\rm Hill} \equiv F_0 + F_2$.

Therefore, if $F_p = \Re e\,[\zeta^{{\rm c}_p}\sum_j f_j\zeta^{2j}]$,
$\sigma$ will have the form $\sigma(\tau)=\Re e\,[\zeta^{{\rm c}_p}
\sum_j \sigma_j\zeta^{2j}]$.
In other words, ${\rm c}_\sigma = \pm {\rm c}_p$, so that the
perturbing potentials $F_1$, $F_3$, $F_5$, $\ldots$ generate source
terms with ${\rm c}_\sigma = \pm 1$, ${\rm c}_\sigma = \pm 3$,
${\rm c}_\sigma = \pm 5$, $\ldots$, respectively.

\section{ Laplace on the equivalence principle}

In the first volume of his {\it Trait\'e de M\'ecanique C\'eleste}
\cite{L2} (presented to the French Academy of Sciences in 1799),
Laplace lists a series of facts suggesting that gravity is
proportional to the masses. This list (which is probably inspired
by a corresponding list in the {\it Principia}, although Laplace
does not mention Newton here) contains Newton's argument that the
motion of satellites would be very sensitive to a violation of the
universality of free fall, but does not quantify it. As far as we
are aware, the only quantitative work of Laplace on this idea
\cite{L1} is contained in the last book of the last volume of the
{\it Trait\'e de M\'ecanique C\'eleste} which was presented to the
Acad\'emie des Sciences on 16 August 1825 \cite{Dic}.

The fact that Laplace was then $76$ year old (he died a year and
a half later on 5 March 1827) may explain why this work of Laplace
contains some strange leaps of reasoning\footnote{ Basically
he mentions that his lunar theory and combined (selected) lunar
and solar parallax data agree to about $1.2$ \% and then goes on to
admit $1/8 = 12.5$ \% as fractional upper limit on the
theory/observation agreement.}. It seems plausible to us that
Laplace, when writing this chapter, was using previous notes of his
which contained more detailed calculations and more consistent
reasonings. Anyway, the aim of this Appendix is to show that the
final limit he quotes on a possible violation of the equivalence
principle,
\begin{equation}
 |{\bar \delta}_{12}| < {1 \over 3410000} \simeq 2.9\times 10^{-7}
 \; , \label{done}
\end{equation}
is a very reasonable (and slightly pessimistic) bound, which can be
derived in a logically clear way using only the information Laplace
had in hand.

Laplace's new idea (compared to Newton) was to use the
``parallactic inequality'' in the longitude of the Moon as a
sensitive test of the equivalence principle\footnote{ Note that as
early as 1753, T.~Mayer had used the theory of the parallactic
inequality to infer the value of the solar parallax \cite{Houz}.}.
``Parallactic inequality'' means the coefficient of the synodic term
$-\sin\tau$ in the expression of the lunar longitude $v$ as a
function of time\footnote{ Beware that Laplace was actually working
with the inverse function: $t=t(v)$.}. We have computed this
coefficient, say $A$, in Appendices~A and B for the Main Problem
(i.e. neglecting eccentricities and inclinations). Its theoretical
expression reads
\begin{equation}
 A^{\rm th} = A_{\rm PAR} + A_{\rm EP}  \; , \label{dtwo}
\end{equation}
where the normal ``parallactic'' contribution reads
\begin{equation}
 A_{\rm PAR} = {15\over 8}(X_2-X_1)\,m{{\tilde a}\over a'}
 \,S_{\rm PAR}(m) \; , \label{dthree}
\end{equation}
while the ``equivalence principle'' one reads
\begin{equation}
 A_{\rm EP} = 3 m\,{\bar \delta}_{12}{a'\over {\tilde a}}
 \,S_{\rm EP}(m) \; . \label{dfour}
\end{equation}
Here, $S_{\rm PAR}(m)=1+{26\over 5}m+\cdots$,
$S_{\rm EP}(m)=1+{16\over 3}m+\cdots$, are slowly
converging series in powers of $m=n'/(n-n')$. In the
third volume of his {\it Trait\'e de M\'ecanique C\'eleste}, Laplace
computes $A_{\rm PAR}$ with particular care, pushing the calculation
to fifth order in $m$ inclusively\footnote{ We have
checked the first orders of his result and found them to agree
with ours. Note that Laplace includes the effect of eccentricities
and inclinations that we neglect.}. He was therefore entitled to
considering that the theoretical error on $A_{\rm PAR}$ was negligible
compared to the observational uncertainties in $A^{\rm obs}$.
Note that $A_{\rm PAR}$ is proportional to the inverse distance to
the Sun, i.e. to the ``solar parallax'' $\pi_S \equiv R_E/a'$, where
$R_E$ is the equatorial radius of the Earth (hence the name
``parallactic inequality''). The result of Laplace can be expressed
as
\begin{equation}
 A_{\rm PAR} \simeq 14.3\, \pi_S \; , \label{dfive}
\end{equation}
where both $A_{\rm PAR}$ and $\pi_S$ are expressed in seconds of
arc\footnote{ Beware that Laplace, in his volume III, uses
(R\'evolution oblige)
``decimal seconds'', i.e. $10^{-6}$ of a right angle.}.

On the observational side, Laplace had in hand both lunar data
and data on the solar parallax. The two phenomenological ``lunar
tables'' he was using (one by Mason and one by Burg) gave for the
``observed'' value of the synodic inequality in longitude,
$A^{\rm obs}_{\rm Mason} = 116."68$ and $A^{\rm obs}_{\rm Burg} =
122."38$ \cite{L3}. As for the solar parallax, many scientific
expeditions had taken advantage of the passages of Venus in front
of the Sun in 1761 and, especially, 1769 to measure $\pi_S$
(see \cite{lal} for a detailed text by a contemporary of Laplace,
or \cite{woolf} for a more exhaustive historical treatment). The
published results ranged between $\pi_S^{\rm obs}=8."43$ (Planman)
and $\pi_S^{\rm obs}=8."80$ (Pingr\'e) \cite{Houz}, \cite{woolf}.
The comparison between the theoretical results
(\ref{dtwo}), (\ref{dfive}) and the observational results on $A$ and
$\pi_S$ gives a value for a possible equivalence-principle-violation
contribution
\begin{equation}
 A_{\rm EP} = A^{\rm obs} - 14.3 \pi_S^{\rm obs}\; . \label{dsix}
\end{equation}

Worst-case limits on $A_{\rm EP}$ are obtained by taking the extreme
values on the right-hand side of (\ref{dsix}) (e.g. $A_{\rm EP}^{\rm
max} = A^{\rm obs}_{\rm Burg} -14.3 \pi_S^{\rm Planman}$)\footnote{
Such a worst-case approach seems appropriate to a
pre-least-squares-law period. Before Gauss' theory of measurement
errors, scientists quoted only ``central values'' for measured
quantities.}. This yields
\begin{equation}
 -9."2 < A_{\rm EP} < 1."8 \; .\label{dseven}
\end{equation}
On the other hand, the theoretical result (\ref{dfour}) [using $
a'/{\tilde a} \simeq 391$ and $S_{\rm EP}(m) \simeq 1.72$
(see Appendix~B)] reads $A_{\rm EP} \simeq 163\, {\bar \delta}_{12}
\simeq 3.36\, {\bar \delta}_{12}\times  10^7$ in seconds of arc,
so that we get the following worst-case bounds on ${\bar \delta}_{12}$
\begin{equation}
 -2.7\times 10^{-7} < {\bar \delta}_{12} < 0.54 \times 10^{-7} \; .
 \label{deight}
\end{equation}
{}From this point of view, the final bound quoted by Laplace,
Eq.~(\ref{done}), seems very reasonable and consistent with the
observational uncertainties in his time. Note, however, that Laplace
never quotes a precise theoretical formula for $A_{\rm EP}$. He only
says (and uses) the fact that the synodic amplitude $A$ is proportional
to its source term in the perturbing function. This neglects the
leading ``$p$ dependence'' of $C_v \propto 2+p$ in
Eq.~(\ref{anineteenb}) which says that $A_{\rm EP}/A_{\rm PAR} \propto
3/5\times S_{\rm EP}(m)/S_{\rm PAR}(m)$. In Laplace's
published analysis the lacking (unfavourable) factor $3/5$ is
effectively
compensated by his overpessimistic estimate of the fractional
uncertainty on $A$: $1/8 = 12.5$ \%\footnote{ As for the subleading
dependence on $p$, i.e. the ratio $S_{\rm EP}(m)/S_{\rm PAR}
(m) \simeq 1.72/1.60 \simeq 1.08$ Laplace's experience with
similar factors in many terms of lunar theory might have suggested
him that he did not need to worry about it.}.

Let us end this section by raising a historical question. Though
Laplace was fully aware of the scientific interest of the bound
(\ref{done}), and of the fact that it was (at the time) more
precise than the bounds obtained from ground tests of the
universality of free fall, his successors in celestial mechanics
seem (as far as we know) to have lost interest in the issue.
However, near the end of the nineteenth century, especially after
the theoretical work of Hansen, Delaunay, Hill and Brown (who
improved the computation of $A^{\rm th}_{\rm PAR}$), and after
many improvements on the observational side, it should have been
possible to obtain more stringent limits on ${\bar \delta}_{12}$.
For instance, Delaunay computed $A^{\rm th}_{\rm PAR}$ to
$m^7$ which corresponds to a truncature error of
$0."3366/125."4201 \simeq 2.7\times 10^{-3}$ \cite{DHR} (for the
principal part of $A$; see also Appendix~B). This is negligible
(when added in quadrature)
compared to the observational error on $\pi_S$ at the time. For
instance,
the laboratory measurements of the velocity of light by Foucault and
Cornu gave values ranging between $\pi_S = 8."834$ and $\pi_S =
8."881$, the passage of Venus across the solar disk in 1874 gave a
range $8."76 - 8."88$ \cite{Houz}, and the recommended value
starting in 1896 was $8."80$. This suggests that a reasonable upper
bound on the uncertainty of $\pi_S$ at that time was $0."08$
(i.e. $0.9$ \%). [By comparison, the modern value is $\pi_S =
8."794148$ \cite{CT95}.] On the other hand, the observational error
on $A^{\rm obs}$ at the time was $< \case1/2(125."46 - 124."70)=
0."38$, i.e. $<0.3$ \% (see \cite{Houz}, p.~533). This has a
negligible effect on the derivation of a bound on ${\bar
\delta}_{12}$\footnote{ Let us note for completeness that the
determinations of the mass of the Moon at the time were accurate
enough to estimate with negligible error the mass ratio factor
$X_2-X_1$ entering $A_{\rm PAR}$.}. Using Eq.~(\ref{dsix}) with
$A_{\rm EP} \simeq 3.36 \,{\bar \delta}_{12}\times 10^7$ seconds of
arc, this leads to the bound $|{\bar \delta}_{12}| < 14.3\, \delta\,
\pi_S^{\rm
obs}/(3.36 \times 10^7) \simeq 3.4\times 10^{-8}$, which is slightly
better than the value obtained by E\"otv\"os in 1890 ($|{\bar
\delta}_{AB}| < 5\times 10^{-8}$ \cite{E1}). We do not wish to
take too seriously such {\it a posteriori} derivations of limits
on ${\bar \delta}_{12}$, but we consider this as an interesting
example of historical eclipse of a deep concept, which has been
rejuvenated, within a new theoretical and observational context,
only in the last decades\footnote{ For completeness, let us note
that Poincar\'e kept alive this concept by mentioning Laplace's
result in one of his popular books \cite{P2}.}.

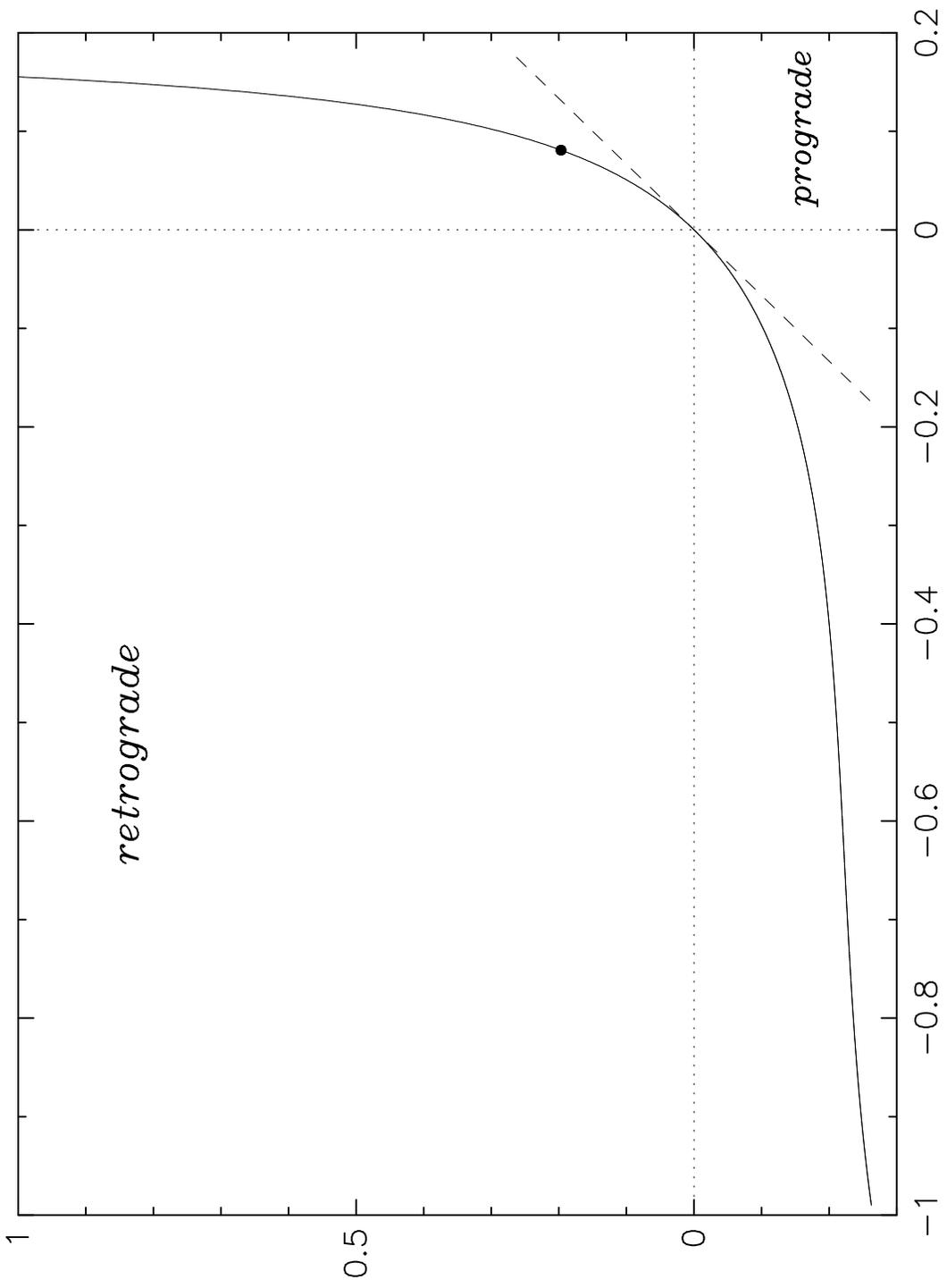
\end{document}